\documentclass[twocolumn,english,superscriptaddress]{revtex4-1}
\usepackage[T1]{fontenc}
\usepackage[latin9]{inputenc}
\usepackage{color}
\usepackage{amsmath}
\usepackage{amssymb}
\usepackage{stmaryrd}
\usepackage{graphicx}
\usepackage{esint}

\makeatletter

\@ifundefined{textcolor}{}{%
 \definecolor{BLACK}{gray}{0}
 \definecolor{WHITE}{gray}{1}
 \definecolor{RED}{rgb}{1,0,0}
 \definecolor{GREEN}{rgb}{0,1,0}
 \definecolor{BLUE}{rgb}{0,0,1}
 \definecolor{CYAN}{cmyk}{1,0,0,0}
 \definecolor{MAGENTA}{cmyk}{0,1,0,0}
 \definecolor{YELLOW}{cmyk}{0,0,1,0}
}

\usepackage{babel}

\usepackage{babel}

\makeatother

\usepackage{babel}
\begin{document}

\title{Superconductivity mediated by quantum critical antiferromagnetic
fluctuations: the rise and fall of hot spots}

\author{Xiaoyu Wang}

\affiliation{School of Physics and Astronomy, University of Minnesota, Minneapolis
55455, USA}

\author{Yoni Schattner}

\affiliation{Department of Condensed Matter Physics, Weizmann Institute of Science,
Rehovot, Israel 76100}

\author{Erez Berg}

\affiliation{Department of Condensed Matter Physics, Weizmann Institute of Science,
Rehovot, Israel 76100}

\author{Rafael M. Fernandes}

\affiliation{School of Physics and Astronomy, University of Minnesota, Minneapolis
55455, USA}
\begin{abstract}
In several unconventional superconductors, the highest superconducting
transition temperature $T_{c}$ is found in a region of the phase
diagram where the antiferromagnetic transition temperature extrapolates
to zero, signaling a putative quantum critical point. The elucidation
of the interplay between these two phenomena \textendash{} high-$T_{c}$
superconductivity and magnetic quantum criticality \textendash{} remains
an important piece of the complex puzzle of unconventional superconductivity.
In this paper, we combine sign-problem-free Quantum Monte Carlo simulations
and field-theoretical analytical calculations to unveil the microscopic
mechanism responsible for the superconducting instability of a general
low-energy model, called spin-fermion model. In this approach, low-energy
electronic states interact with each other via the exchange of quantum
critical magnetic fluctuations. We find that even in the regime of
moderately strong interactions, both the superconducting transition
temperature and the pairing susceptibility are governed not by the
properties of the entire Fermi surface, but instead by the properties
of small portions of the Fermi surface called hot spots. Moreover,
$T_{c}$ increases with increasing interaction strength, until it
starts to saturate at the crossover from hot-spots dominated to Fermi-surface
dominated pairing. Our work provides not only invaluable insights
into the system parameters that most strongly affect $T_{c}$, but
also important benchmarks to assess the origin of superconductivity
in both microscopic models and actual materials. 
\end{abstract}
\maketitle

\section{Introduction}

\begin{figure}
\begin{centering}
\includegraphics[width=1\columnwidth]{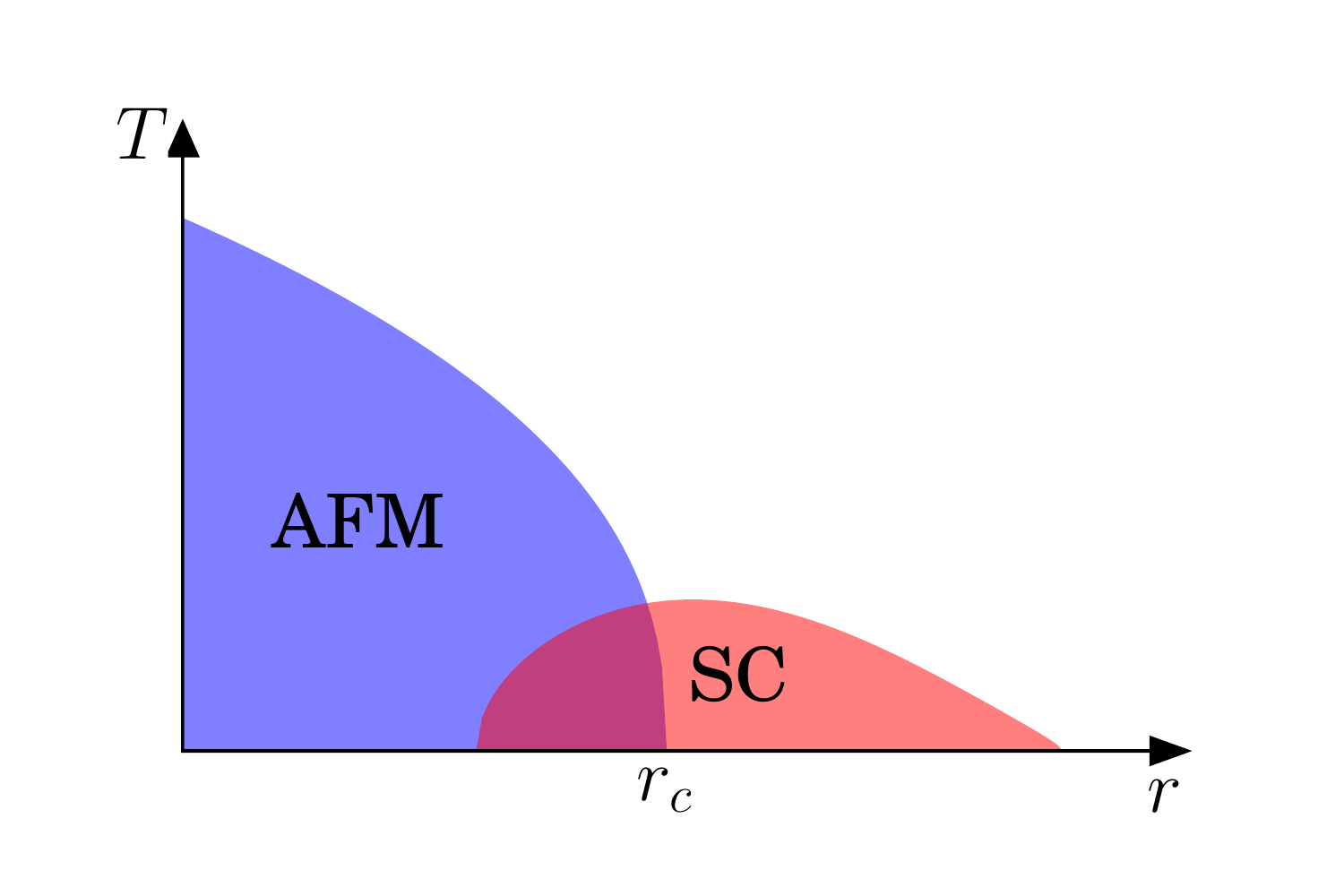} 
\par\end{centering}
\caption{Schematic phase diagram of the spin-fermion model. The antiferromagnetic
(AFM) transition temperature is suppressed to zero at $r=r_{c}$,
giving rise to a quantum critical point. According to the results
of Ref. \cite{Schattner16} for the spin-fermion model, a superconducting
(SC) dome then appears, hiding the antiferromagnetic quantum critical
point. The maximum $T_{c}$ is found very close to $r=r_{c}$. \label{fig_schematic}}
\end{figure}

In the two known families of high-temperature superconductors \textendash{}
the copper-based and the iron-based materials \textendash{} the superconducting
(SC) state is observed in close proximity to an antiferromagnetic
(AFM) state \cite{Varma86,Scalapino86,Pines92,Hirschfeld11}. In the
particular cases of iron pnictides and electron-doped cuprates, the
highest SC transition temperature $T_{c}$ takes place in the vicinity
of a putative antiferromagnetic quantum critical point (QCP) \cite{QCP_cuprates1,QCP_cuprates3,QCP_pnictides2},
i.e. a continuous AFM phase transition that occurs at zero temperature
(see Fig. \ref{fig_schematic}). Although direct detection of such
a QCP is difficult, some of its manifestations at non-zero temperatures,
such as a nearly-diverging magnetic correlation length, are experimentally
observed \cite{QCP_cuprates2,QCP_pnictides1}. These observations
led to the proposal that quantum critical AFM fluctuations may provide
the glue binding the Cooper pairs in an unconventional SC state \cite{Chubukov03,Tremblay03,Sachdev10,Pepin13,Millis92,Chubukov16},
be it a nodal $d$-wave state, as in the case of the cuprates, or
a nodeless $s^{+-}$-wave state, as in the case of the iron pnictides.

The reasoning behind this theoretical proposal can be understood from
a straightforward extension of the conventional weak-coupling BCS
theory for phonon-mediated $s$-wave superconductors. In contrast
to the electron-phonon coupling, which causes an attractive pairing
interaction that does not depend on momentum, AFM fluctuations generate
a repulsive pairing interaction strongly peaked at the momentum corresponding
to the AFM wave-vector $\mathbf{Q}$ \cite{Varma86,Scalapino86,Pines92,Hirschfeld11}.
In this case, the BCS gap equations only admit a solution if the gap
function $\Delta\left(\mathbf{k}\right)$ changes its sign when the
momentum is translated by $\mathbf{Q}$, i.e. $\Delta\left(\mathbf{k}+\mathbf{Q}\right)\propto-\Delta\left(\mathbf{k}\right)$.
As a result, depending on the Fermi surface geometry and on the wave-vector
$\mathbf{Q}$, different types of SC states are favored. While a $d$-wave
state is obtained for a large Fermi surface and $\mathbf{Q}=\left(\pi,\pi\right)$,
an $s^{+-}$-wave state arises for small Fermi pockets separated by
$\mathbf{Q}=\left(\pi,0\right)/(0,\pi)$. Despite its appeal, such
a weak-coupling BCS-like approach is not appropriate to describe these
systems, since the proximity to a QCP renders the interactions strong
and, on top of that, clouds the very concept of quasi-particles, which
is a key property of a Fermi liquid \cite{Lohneysen07}.

Thus, while there is little question that AFM fluctuations can promote
an unconventional SC state, the elucidation of\emph{ }the microscopic
mechanisms involved remain a major challenge. Addressing this issue
is important not only to assess the relevance of quantum critical
pairing to high-$T_{c}$ materials, but also to establish which of
the many system parameters should be ideally optimized to enhance
$T_{c}$. To answer these important questions, microscopic models
that are expected to display AFM and SC ground states have been widely
studied, most notably the Hubbard model \cite{Hubbard1,Hubbard2,Hubbard3,Hubbard4}.
Alternatively, in the hope to elucidate universal features of quantum
critical pairing, many works have focused on a general low-energy
model in which the fermions associated with the low-energy electronic
states interact with each other by exchanging magnetic fluctuations,
which in turn arise from high-energy states \textendash{} this is
the so-called spin-fermion model \cite{Chubukov03,Sachdev10}. Because
these fluctuations are peaked at the AFM ordering vector $\mathbf{Q}$,
not all low-energy states are equally affected by this interaction.
More specifically, only states near the hot spots \textendash{} special
points on the Fermi surface that are displaced from each other by
the AFM ordering vector $\mathbf{Q}$ \textendash{} can exchange AFM
fluctuations while remaining near the Fermi level. This property lends
support to the idea that the hot spots may play a primary role in
driving the superconducting transition.

However, despite intense research activity in this front, the extent
to which hot-spots properties govern the SC instability remains a
hotly debated issue. One of the reasons is the difficulty in developing
a controlled strong-coupling theory for the spin-fermion model, which
is ultimately related to the absence of a natural small parameter
in the problem \cite{SSLee09,Sachdev10,Senthil10,Raghu15}. This situation
is to be contrasted with the phonon-mediated pairing problem, where
the clear separation between electronic and lattice energy scales
ensures the existence of a controlled diagrammatic expansion \textendash{}
the celebrated Eliashberg theory.

In this paper, we combine extensive Quantum Monte Carlo (QMC) simulations
and analytical calculations to shed light on this problem. Our starting
point is the two-band version of the two-dimensional spin-fermion
model, in which the AFM fluctuations mediate interactions between
electrons from two different bands. The choice of a two-band model
is essential, because it does not suffer from the infamous sign-problem
generally present in QMC simulations \cite{Berg12}\textcolor{black}{.
While a recent study has established the existence of a SC dome peaked
at the AFM quantum critical point of this model \cite{Schattner16},
similarly to Fig. \ref{fig_schematic}, in this paper our goal is
to elucidate the microscopic mechanism responsible for this SC state.
}Unveiling the pairing mechanism encoded in the spin-fermion model
is fundamental to advance our understanding of the general problem
of superconductivity in quantum critical systems for several reasons.
First, the sign-problem-free QMC algorithm only works for the rather
artifical two-band model. Establishing the solution of this two-band
spin-fermion model, where the unbiased sign-problem-free QMC approach
offers a unique benchmark for analytical approximations, is the most
promising way to generalize the results to other types of band structures.
Second, being a low-energy model, the main relevance of the spin-fermion
model to the ongoing effort to search for higher $T_{c}$ materials
is to provide robust trends for how changes in the various system
parameters affect $T_{c}$ \textendash{} e.g. is the density of states
at the Fermi level more important than the properties of the hot spots?
Third, the spin-fermion model is one among several models that have
been proposed to understand high-$T_{c}$ systems. Without knowing
the precise predictions of this model, it is very hard to rule out
or confirm that the physics encoded in the spin-fermion model is relevant
to the real systems.

In this paper, our general goal is to establish the general solution
of the spin-fermion model by a detailed comparison between numerics
and analytics. To achieve this goal, we study a family of band dispersions
that interpolate between closed nearly-nested Fermi pockets to open
Fermi surfaces, passing through a van Hove singularity, where the
density of states is strongly peaked. This non-trivial dependence
of the density of states on the band dispersion allows us to separate
phenomena associated with the Fermi surface as a whole and with the
hot spots only. Tuning the system to its AFM quantum critical point,
we extract from our QMC results both the superconducting transition
temperature $T_{c}$ \textendash{} which in our two-dimensional system
is a Berezinski-Kosterlitz-Thouless transition \textendash{} and the
temperature dependence of the pairing susceptibility, $\chi_{\mathrm{pair}}$.
Similarly to previous works \cite{Berg12,Schattner16,DHLee16_2},
we find that the favored SC state is the one in which the gap function
changes sign from one band to the other \textendash{} in qualitative
agreement with the weak-coupling arguments given above. Our main results,
however, are on the dependence of $T_{c}$ and $\chi_{\mathrm{pair}}$
on the band dispersion parameters. Surprisingly, we find that $T_{c}$
is not sensitive to the density of states $N_{f}$, which displays
a sharp enhancement near the van Hove singularity. Instead, even when
the interaction strength is comparable to the bandwidth, $T_{c}$
is found to depend only on the angle between the Fermi velocities
of a pair of hot spots, $\sin\theta_{\mathrm{hs}}$, via:

\begin{equation}
T_{c}=A_{c}\lambda^{2}\sin\theta_{\mathrm{hs}}\label{T_BKT}
\end{equation}
where $\lambda$ is the interaction parameter that couples magnetic
and electronic degrees of freedom, and $A_{c}$ is a universal constant
independent of the band dispersion. As for the pairing susceptibility,
we show that the QMC data for all band dispersions collapse onto a
single curve given by:

\begin{equation}
\chi_{\mathrm{pair}}\left(T\right)=A_{\mathrm{pair}}f_{\mathrm{pair}}\left(\frac{T}{T_{c}}\right)\label{chi_pair}
\end{equation}
where $f_{\mathrm{pair}}\left(\frac{T}{T_{c}}\right)$ is a universal
function that does not depend on the band dispersion, whereas $A_{\mathrm{pair}}$
is a constant that depends weakly on the band dispersion. Eqs. (\ref{T_BKT})
and (\ref{chi_pair}) are the main results of our paper, establishing
that the hot-spots properties govern not only the SC transition temperature,
but also the temperature dependence of the SC fluctuations. To understand
these results, we analytically study the spin-fermion model via a
hot-spots Eliashberg approximation introduced in previous works for
the one-band model \cite{Chubukov03,Chubukov16}. Basically, this
approximation consists of assuming that the magnetic degrees of freedom
are much slower than the electronic ones, and that the hot spots govern
the critical properties of the system. Despite being formally uncontrolled,
this approximation not only gives the same functional dependence of
the SC transition temperature on the spin-fermion parameters of Eq.
(\ref{T_BKT}), but it also captures very well the universal function
$f_{\mathrm{pair}}\left(x\right)$ obtained from the QMC results.

An immediate consequence of Eq. (\ref{T_BKT}) is that $T_{c}$ would
not have an upper limit upon increasing the interaction $\lambda$.
We find, however, that when $\lambda^{2}$ becomes larger than the
electronic bandwidth, $T_{c}$ stops increasing and nearly saturates
to a value of the order of a few percent of the electronic bandwidth.
Combined with our analytical investigation of the spin-fermion model,
we attribute this behavior to the whole Fermi surface behaving as
a ``large hot-spot,'' and to the failure of the hot-spots-only approximation
\cite{Abanov08}. Therefore, our results indicate that, within the
spin-fermion model, the largest possible value of $T_{c}$ does not
depend on the interaction strength, and is first achieved at the crossover
between hot-spots dominated and Fermi-surface dominated pairing.

\begin{figure}
\begin{centering}
\includegraphics[width=0.9\columnwidth]{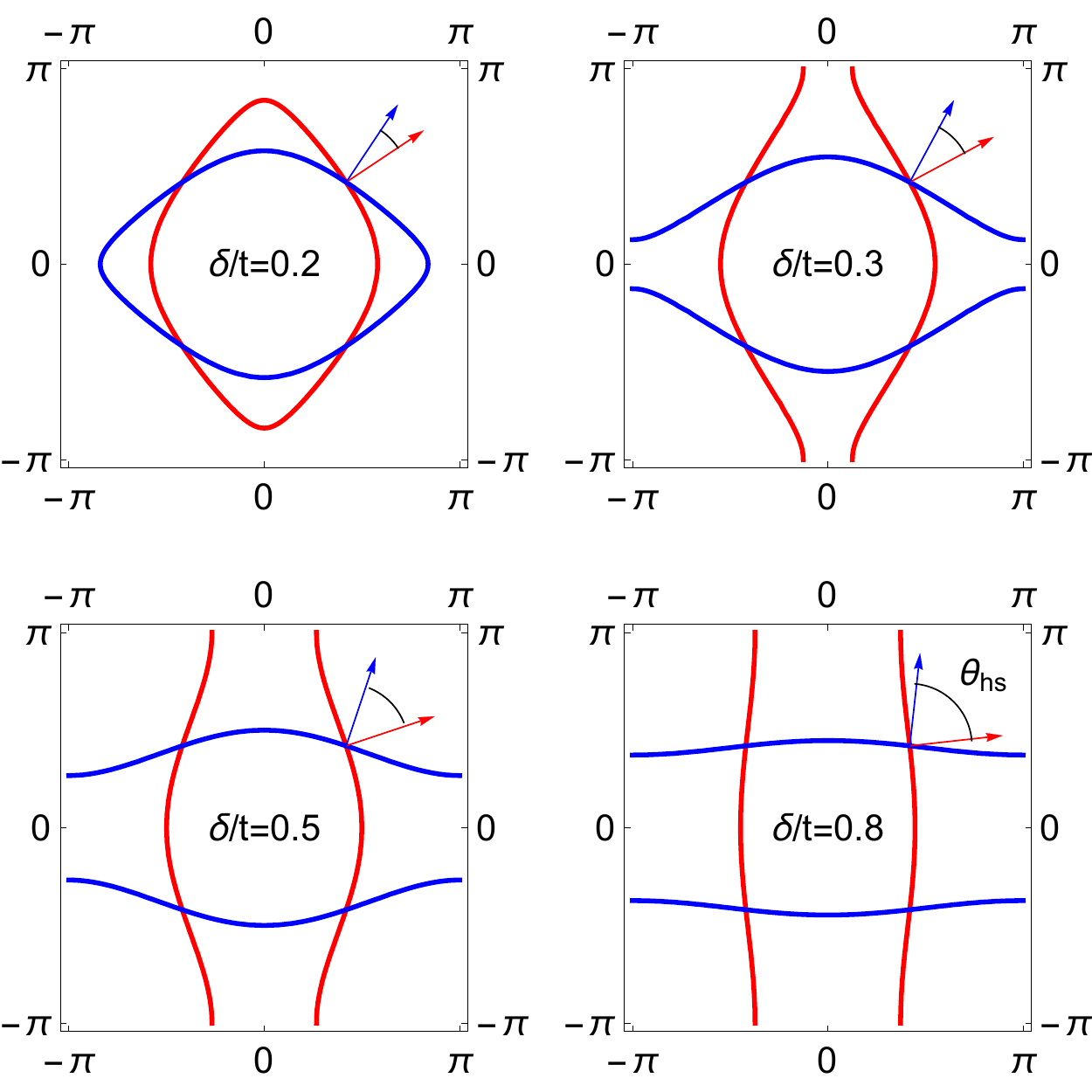} 
\par\end{centering}
\caption{The two-band spin-fermion model. Fermi surfaces corresponding to the
two bands (red and blue curves) in the first Brillouin zone, for different
values of $\delta/t$. One of the bands (blue) is displaced by the
AFM wave-vector $\mathbf{Q=\left(\pi,\pi\right)}$, which makes both
Fermi surfaces appear concentric. In this representation, a pair of
hot spots, defined by $\varepsilon_{c,\mathbf{k}_{\mathrm{hs}}}=\varepsilon_{d,\mathbf{k}_{\mathrm{hs}}+\mathbf{Q}}=0$,
correspond to the points at which the two Fermi surfaces overlap.
For the system parameters used here, the hot spots are always along
the diagonals of the Brillouin zone. By changing the parameter $\delta/t$,
the system interpolates between closed nearly-nested Fermi surfaces
($\delta/t<1/4$) and open Fermi surfaces ($\delta/t>1/4$), crossing
a van Hove singularity at $\delta/t=1/4$. The angle $\theta_{\mathrm{hs}}$
between the Fermi velocities of a pair of hot spots (red and blue
arrows) increases as function of $\delta/t$ (note that one of the
Fermi velocities has been multiplied by $-1$ for clarity purposes).
\label{fig_bands}}
\end{figure}

\section{The spin-fermion model}

The spin-fermion model is a low-energy model widely employed to study
universal properties of pairing mediated by AFM fluctuations \cite{Chubukov03,Sachdev10,Berg12}.
It describes low-energy electronic degrees of freedom interacting
with magnetic fluctuations that arise from high-energy degrees of
freedom. In this work, we consider a two-dimensional model with two
independent bands, yielding the following non-interacting Hamiltonian:

\begin{equation}
\mathcal{H}_{0}=\sum_{\mathbf{k}\alpha}\varepsilon_{c,\mathbf{k}}c_{\mathbf{k}\alpha}^{\dagger}c_{\mathbf{k}\alpha}^{\phantom{\dagger}}+\sum_{\mathbf{k},\alpha}\varepsilon_{d,\mathbf{k}}d_{\mathbf{k}\alpha}^{\dagger}d_{\mathbf{k}\alpha}^{\phantom{\dagger}}\,.\label{H0}
\end{equation}
Here, the operator $c_{\mathbf{k}\alpha}^{\dagger}$ creates an electron
with momentum $\mathbf{k}$ and spin $\alpha$ at band $c$. The centers
of the two bands are displaced from each other by the AFM ordering
vector $\mathbf{Q}=\left(\pi,\pi\right)$, and the dispersions are
given by

\begin{align}
\varepsilon_{c,\mathbf{k}} & =\mu-2(t+\delta)\cos k_{x}-2(t-\delta)\cos k_{y}\nonumber \\
\varepsilon_{d,\mathbf{k}+\mathbf{Q}} & =-\mu+2(t-\delta)\cos k_{x}+2(t+\delta)\cos k_{y}\,,\label{bands}
\end{align}
where $t$ is the hopping parameter, $\mu$ is the chemical potential,
and momentum is measured in units of the inverse lattice constant
$1/a$. Note that this model is symmetric under the combination of
a $\pi/2$ rotation, a particle-hole transformation, and the exchange
of the two bands. Hereafter, we set $\mu=t$. By changing the parameter
$\delta$, the band dispersions interpolate between two closed nearly-nested
Fermi pockets ($\delta<t/4$) and two open Fermi surfaces ($\delta>t/4$),
see Fig. \ref{fig_bands}. For $\delta=t/4$, the band dispersion
has a saddle point at the Fermi level, implying the existence of a
van Hove singularity, which is characterized by a diverging density
of states, $N_{f}$.

In the spin-fermion model the electrons interact with each other only
via the exchange of magnetic fluctuations. As a result, the interaction
action is given by:

\begin{equation}
S_{\mathrm{int}}=\lambda\sum_{j}\int_{\tau}\mathbf{M}_{j}\mathrm{e}^{i\mathbf{Q}\cdot\mathbf{x}_{j}}\cdot\left(c_{j,\alpha}^{\dagger}\boldsymbol{\sigma}_{\alpha\beta}d_{j,\beta}+\mathrm{h.c.}\right)\,.\label{H_int}
\end{equation}
Here, $j$ denotes lattice sites, $\tau$ is the imaginary time, $\lambda$
is the (Yukawa) coupling constant describing the interaction between
electrons and magnetic fluctuations, $\boldsymbol{\sigma}$ are Pauli
matrices, and $\mathbf{M}$ is the bosonic field associated with magnetic
order with wave-vector $\mathbf{Q}$. The spectrum of magnetic fluctuations
is determined by the magnetic action, which in turn arises from high-energy
electronic degrees of freedom:

\begin{equation}
S_{\mathrm{mag}}=\frac{1}{2}\int_{\mathbf{x},\tau}\left[\frac{1}{v_{s}^{2}}\left(\partial_{\tau}\mathbf{M}\right)^{2}+\left(\boldsymbol{\nabla}\mathbf{M}\right)^{2}+rM^{2}+\frac{u}{2}\,M^{4}\right]\label{S_mag}
\end{equation}

In this expression, $r$ is a tuning parameter that tunes the system
through the magnetic quantum critical point, $u=1/(2t)>0$ is a parameter
penalizing strong amplitude fluctuations, and $v_{s}=4t$ is the spin-wave
velocity. Note that, in our notation, $\lambda^{2}$ has dimensions
of energy. If $\lambda$ was zero, Eq. (\ref{S_mag}) would describe
a magnetic ordered phase that, at $T=0$, undergoes a second-order
quantum phase transition to a paramagnetic state at $r=r_{c}$ (see
Fig. \ref{fig_schematic}). The coupling to the electrons not only
shifts the value of $r_{c}$, but it also promotes new electronic
ordered phases, most notably superconductivity. Additional details
about the spin-fermion model are given in Appedix A.

\section{Sign-problem-free Quantum Monte Carlo simulations}

Eqs. (\ref{H0}), (\ref{H_int}), and (\ref{S_mag}) define the two-band
spin-fermion model. Because the total fermionic action $S_{0}+S_{\mathrm{int}}$
commutes with an anti-unitary operator for every configuration of
$\mathbf{M}$, all eigenvalues of the fermionic determinant are complex-conjugate
pairs, implying that determinant QMC simulations do not suffer from
the sign-problem \cite{Berg12}. Here, $S_{0}$ is the non-interacting
action associated with $\mathcal{H}_{0}$ in Eq. (\ref{H0}). Previous
QMC studies have shown conclusively that, in this type of models,
the sign-changing SC pairing susceptibility is strongly enhanced near
the magnetic QCP \cite{Berg12,DHLee16_2,Schattner16}. Because the
system is two-dimensional, at finite temperatures only quasi-long-range
SC order is stabilized, which happens below the Berezinskii-Kosterlitz-Thouless
(BKT) transition temperature $T_{c}$. The latter was shown to be
maximum very close to the putative quantum critical point $r=r_{c}$
\cite{Schattner16}. More recently, similar sign-problem-free QMC
approaches have been used to study charge fluctuations near an AFM-QCP
and the onset of SC near a nematic QCP \cite{Dumitrescu15,DHLee16_1,DHLee16_2,Schattner15}.

Here, our goal is to establish which band structure parameters determine
$T_{c}$ and $\chi_{\mathrm{pair}}$, in order to shed light on the
microscopic mechanism by which quantum critical AFM fluctuations promote
superconductivity. Our procedure is the following: for a given band
dispersion, labeled by $\delta/t$, we first determine the approximate
location of the AFM quantum critical point $r_{c}$ by analyzing both
$\left\langle \mathbf{M}^{2}\right\rangle $ and the Binder cumulant.
To save computational time, we consider an easy-plane AFM order parameter,
restricting $\mathbf{M}$ to lie in the XY plane. We verify that the
system is in the magnetically disordered state and very close to the
QCP by computing the renormalized mass term of the magnetic propagator.
Note that for the system to be in a quantum critical regime, it is
enough that the magnetic mass term be much smaller than $\pi T_{c}/\gamma$,
where $\gamma$ is the Landau damping. As long as this condition is
satisfied, even if at $T=0$ the AFM transition becomes weakly first-order,
the system's behavior at finite temperatures would still be nearly
indistinguishable from a quantum critical one. The static pairing
susceptibility in the sign-changing SC channel, $\chi_{\mathrm{pair}}$,
is obtained by direct computation of the pair correlation function,
while the superfluid density $\rho_{s}$ is obtained from the current-current
correlation function\emph{.} We study square lattices of sizes $L=8$,
$L=10$, $L=12$, and $L=14$. Spurious finite size effects are diminished
by threading a fictitious magnetic flux quantum through the system.
Technical details of the QMC implementation are similar to those in
Ref. \cite{Schattner16}, and are summarized in Appendix B.

\begin{figure}
\begin{centering}
\includegraphics[width=0.9\columnwidth]{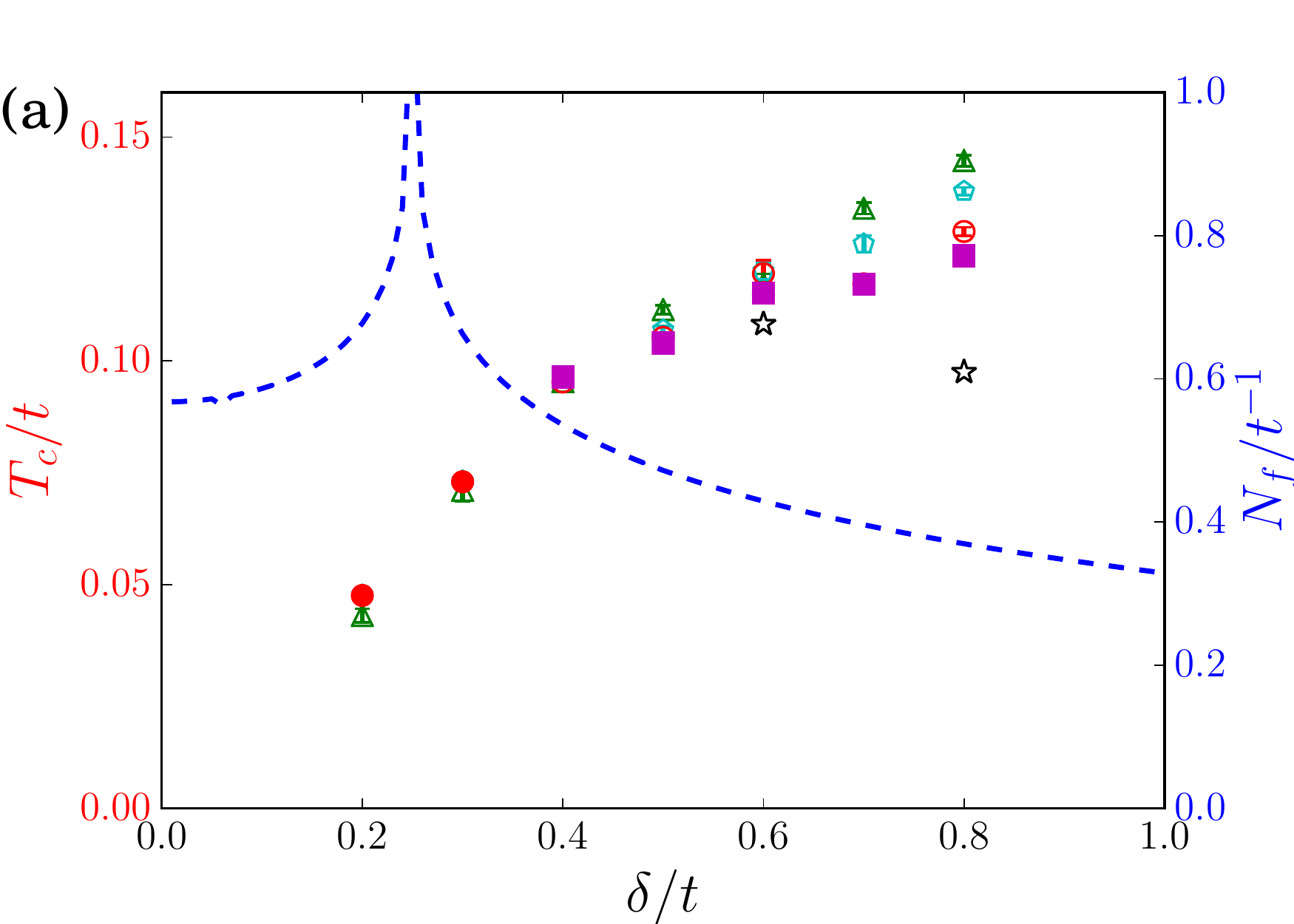} 
\par\end{centering}
\begin{centering}
\includegraphics[width=0.9\columnwidth]{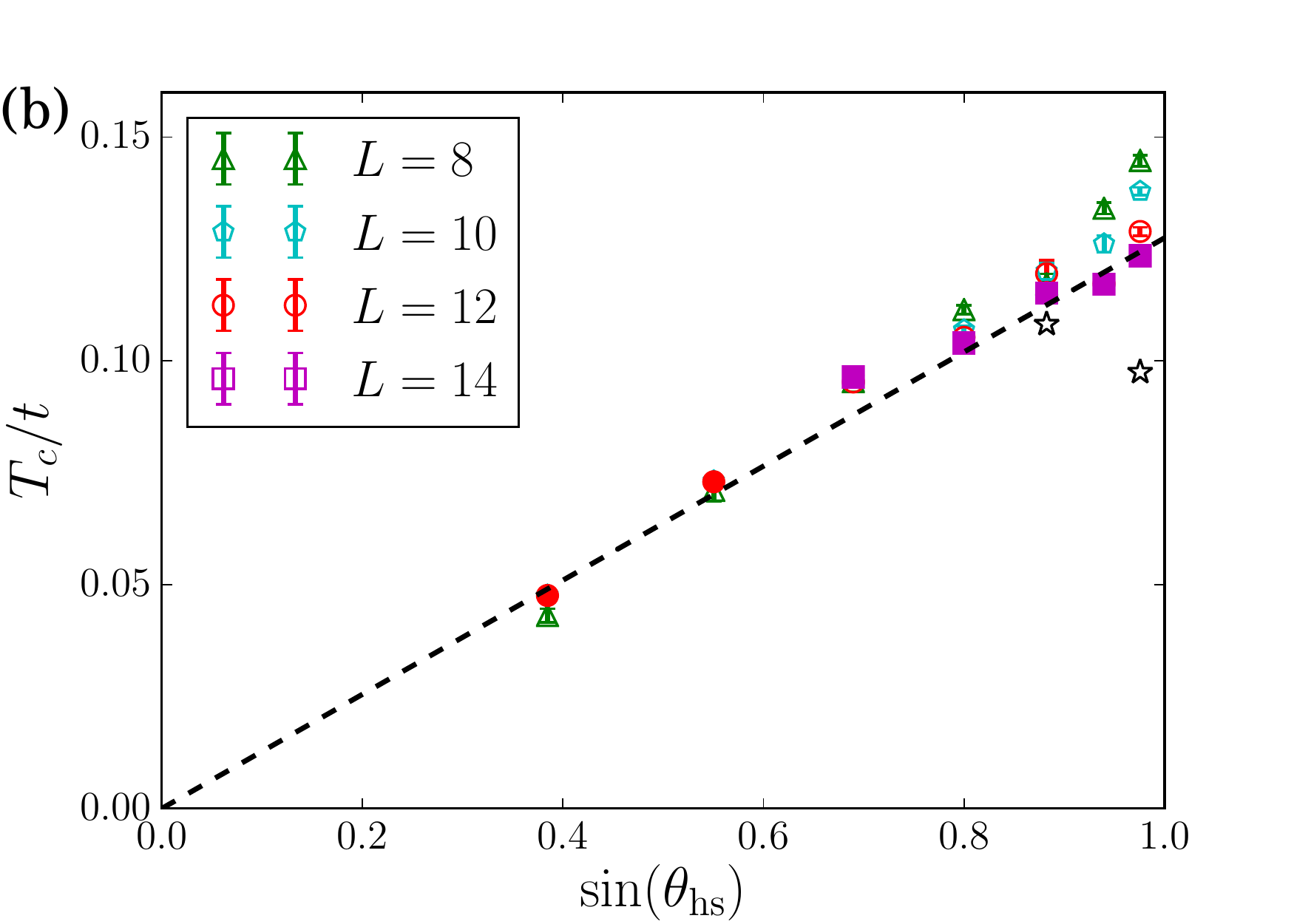} 
\par\end{centering}
\caption{The superconducting transition temperature $T_{c}$ at the QCP for
different band dispersion parameters. (a) The QMC results for $T_{c}$
and the calculated density of states $N_{f}$ (calculated directly
from the band dispersions) as function of the band dispersion parameter
$\delta/t$ (see Fig. \ref{fig_bands}). We associate a transition
temperature $T_{c}\left(L\right)$ to the temperature at which the
BKT condition is met for a system of size $L$, and denote $T_{c}\left(L_{\mathrm{max}}\right)$
by filled symbols. Analysis of finite-size effects reveals that for
most values of $\delta/t$, $T_{c}\left(L_{\mathrm{max}}\right)$
is a very good estimate for the thermodynamic-limit value $T_{c}$.
For the systems in which $T_{c}\left(L\right)$ does not fully converge,
namely $\delta/t=0.6$ and $\delta/t=0.8$, $T_{c}(L_{\mathrm{max}})$
are upper bound values for $T_{c}$, whereas the stars are lower bound
values on $T_{c}$. Note the enhanced $N_{f}$ at the van Hove singularity
point $\delta/t=1/4$. (b) The linear relationship between $T_{\mathrm{c}}$
and $\sin\theta_{\mathrm{hs}}$, where $\theta_{\mathrm{hs}}$ is
the angle between the two Fermi velocities of a pair of hot spots,
calculated directly from the band dispersions. \label{fig_TBKT}}
\end{figure}

For each system size $L$, we associate a transition temperature $T_{c}\left(L\right)$
to the temperature at which the BKT condition is met, $\rho_{s}=2T_{c}/\pi$.
In Fig. \ref{fig_TBKT}, we show the behavior of $T_{c}\left(L\right)$
at the AFM-QCP as function of the parameter $\delta/t$ introduced
in Eq. (\ref{bands}) for a moderately strong interaction parameter
$\lambda^{2}=8t$. For most band dispersion parameters, $T_{c}\left(L\right)$
of the two largest system sizes are coincident within the QMC statistical
error bars. In these cases, our best estimate for the thermodynamic
value of $T_{c}\equiv T_{c}\left(L\rightarrow\infty\right)$ is the
value corresponding to the largest system size, $T_{c}\left(L_{\mathrm{max}}\right)$
(filled symbols in the figure). For the band dispersion parameters
in which $T_{c}$ does not seem to fully converge with system size,
namely $\delta/t=0.6$ and $\delta/t=0.8$, $T_{c}\left(L_{\mathrm{max}}\right)$
should be understood as an upper bound on $T_{c}$. In these cases,
we also provide a lower bound on $T_{c}$, represented by stars in
the figure (see Appendix B for more details of this procedure). Clearly,
the finite size effects seem to affect mostly the band dispersion
with $\delta/t=0.8$, which has a more pronounced one-dimensional
character, as shown in Fig. \ref{fig_bands}d. Interestingly, analytical
studies of the spin-fermion model suggested a strong competition between
SC and charge order for quasi-one-dimensional band dispersions \cite{Sachdev10,Pepin13}.
Whether this is related to the stronger finite size effects observed
for $\delta/t=0.8$ is an interesting topic for future investigation.

Surprisingly, Fig. \ref{fig_TBKT} reveals that $T_{c}$ is not sensitive
to the non-interacting density of states $N_{f}$, which diverges
at the van Hove singularity at $\delta/t=0.25$, as shown in the same
figure (as shown in the supplementary material, even for our finite
systems, $N_{f}$ is also peaked at the van Hove singularity). Instead,
we find that $T_{c}$ increases linearly with $\sin\theta_{\mathrm{hs}}$,
where $\theta_{\mathrm{hs}}$ is the angle between the non-interacting
Fermi velocities of a hot-spot pair (see Fig. \ref{fig_bands}). In
contrast to $N_{f}$, which varies non-monotonically as function of
$\delta/t$, $\sin\theta_{\mathrm{hs}}$ changes monotonically according
to $\sin\theta_{\mathrm{hs}}=\frac{2\left(\delta/t\right)}{1+\left(\delta/t\right)^{2}}$.

\begin{figure}
\begin{centering}
\includegraphics[width=0.9\columnwidth]{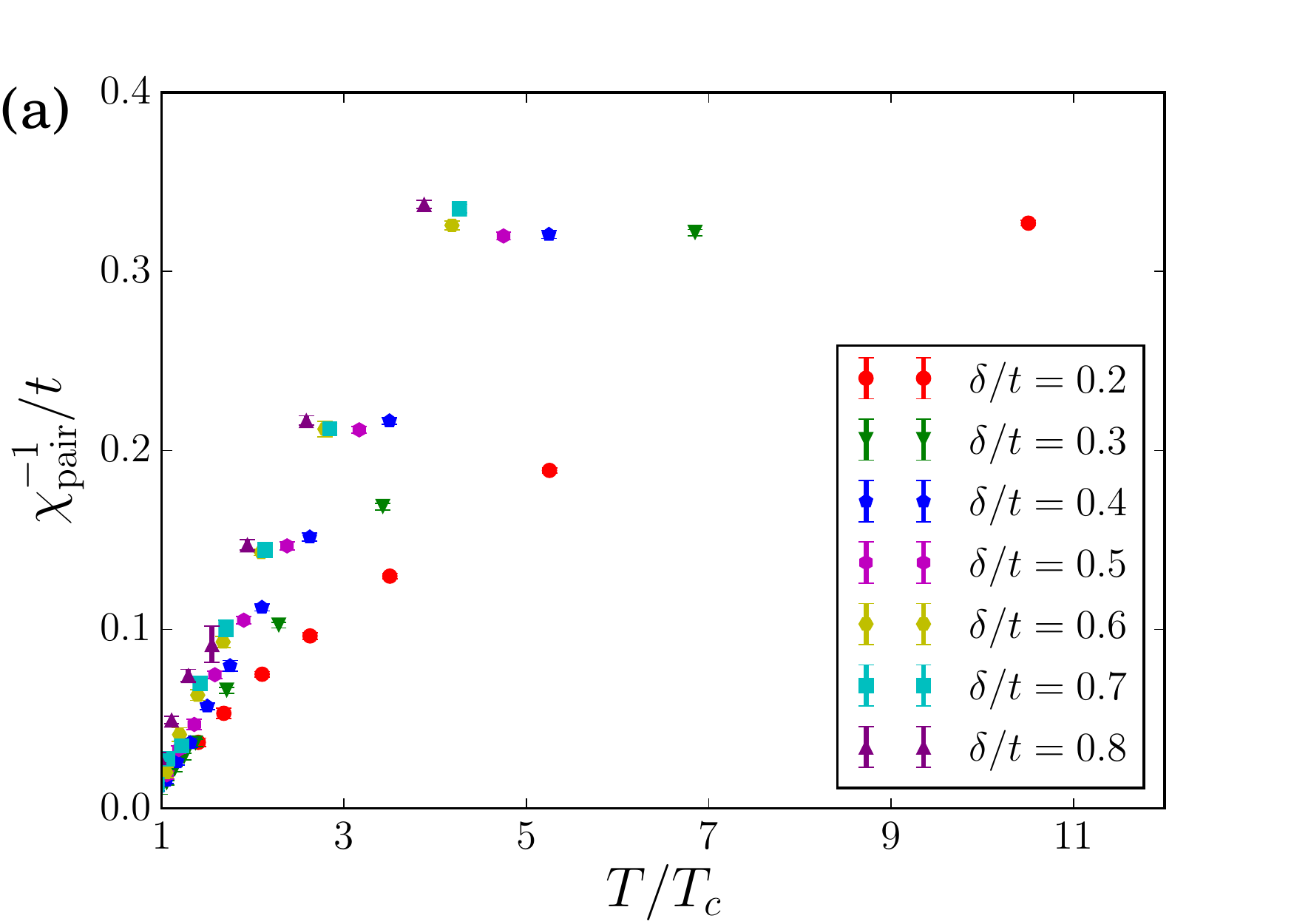} 
\par\end{centering}
\begin{centering}
\includegraphics[width=0.9\columnwidth]{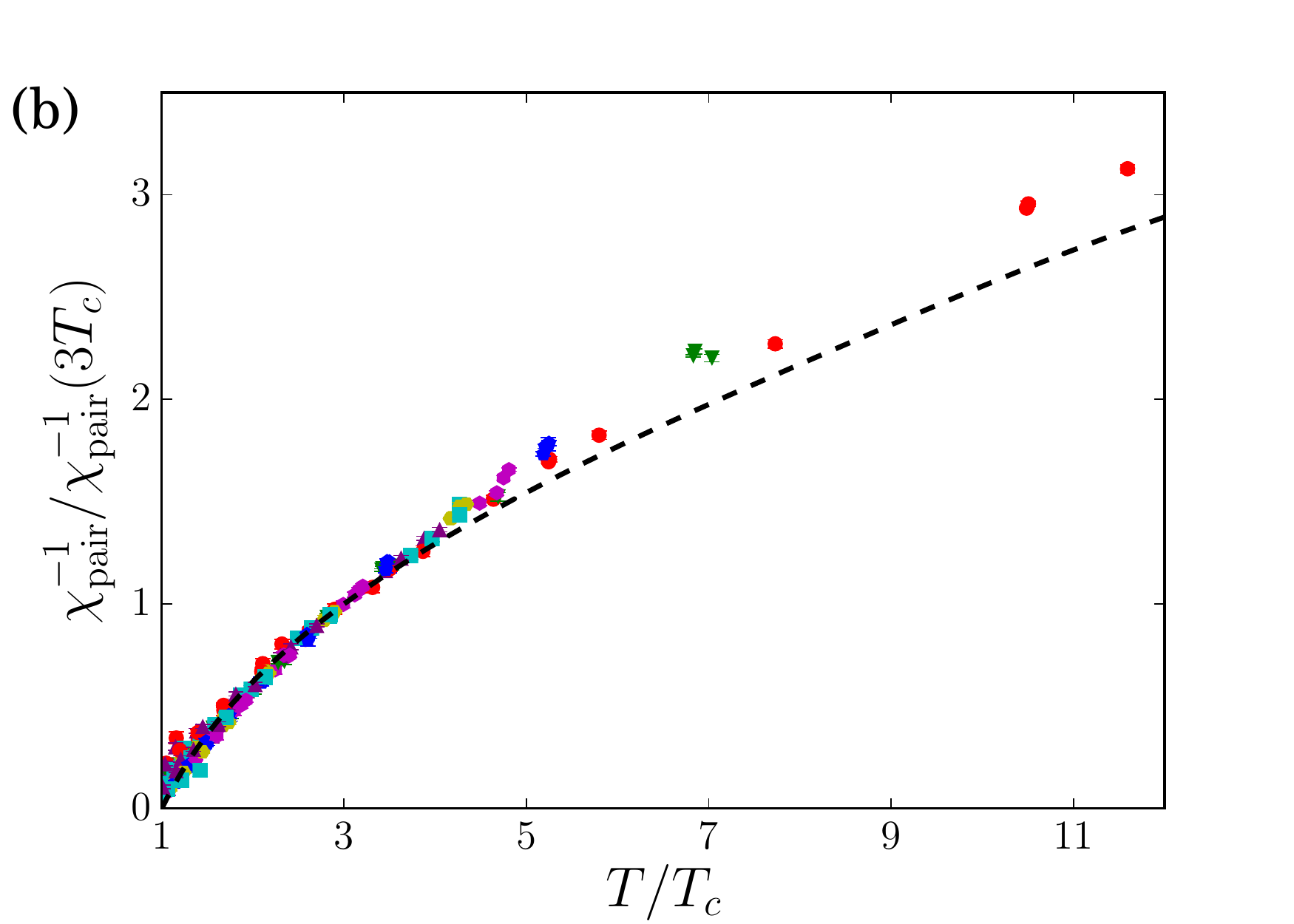} 
\par\end{centering}
\begin{centering}
\includegraphics[width=0.9\columnwidth]{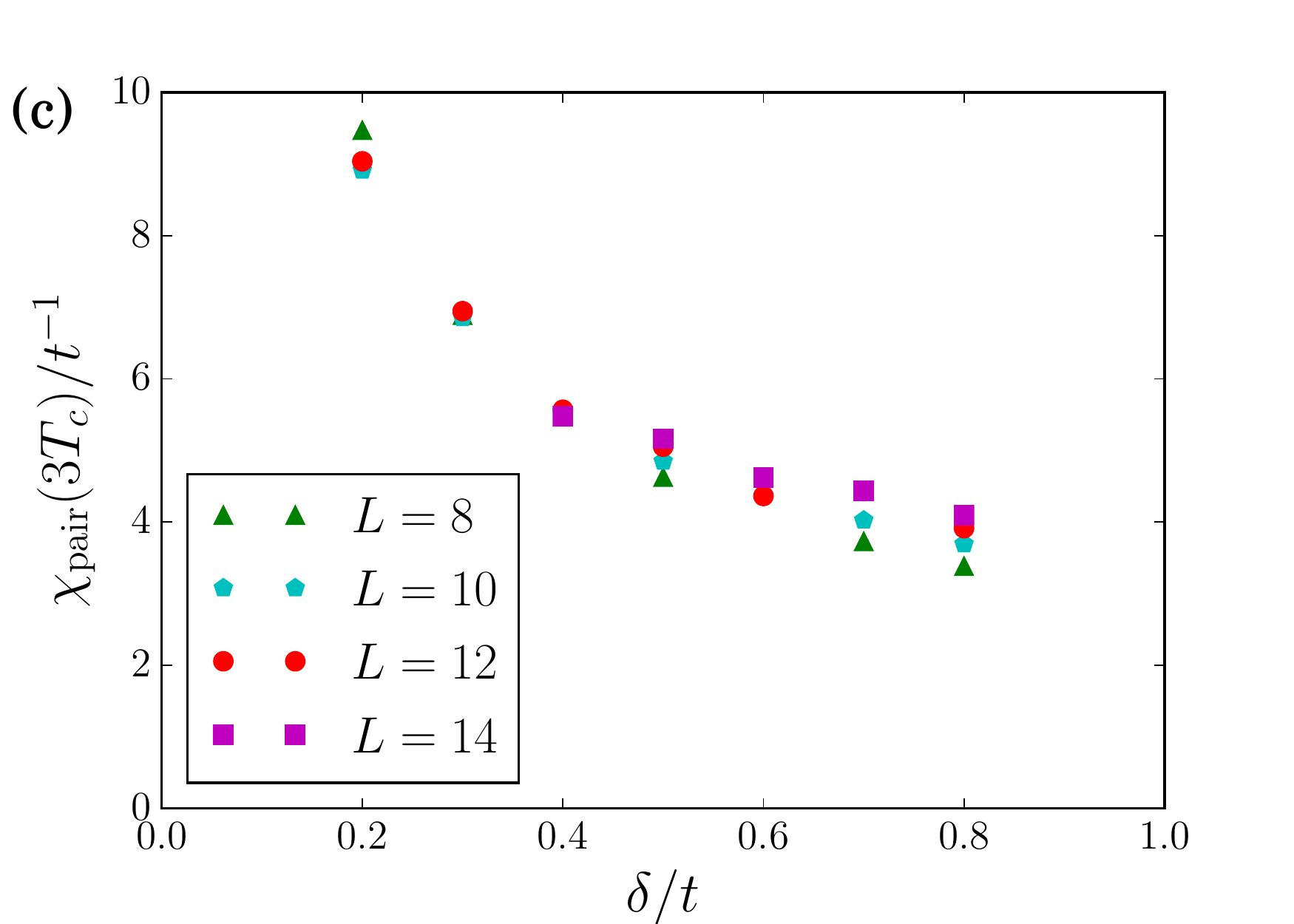} 
\par\end{centering}
\caption{Universal temperature dependence of the pairing susceptibility $\chi_{\mathrm{pair}}$
at the QCP. (a) Temperature dependence of $\chi_{\mathrm{pair}}^{-1}$
extracted from QMC simulations for all band dispersion parameters
$\delta/t$. The system size is $L=12$. (b) Collapse of the scaled
$\chi_{\mathrm{pair}}^{-1}(T)/\chi_{\mathrm{pair}}^{-1}(3T_{c})$
as function of $T/T_{c}$ for all values of $\delta/t$ and all system
sizes $L$. For each value of $L$, we used the corresponding $T_{c}(L)$.
The black dashed curve is the analytical function $f_{\mathrm{pair}}^{(\mathrm{hs})}(T/T_{c})/f_{\mathrm{pair}}^{(\mathrm{hs})}(3)$
obtained from the hot-spots Eliashberg approximation of the spin-fermion
model. (c) The behavior of the QMC-extracted pre-factor $A_{\mathrm{pair}}\propto\chi_{\mathrm{pair}}(3T_{c})$
of Eq. (\ref{chi_pair}) as function of $\delta/t$. \label{fig_susceptibility}}
\end{figure}

The results shown in Fig. \ref{fig_TBKT} imply that the SC transition
is rather insensitive to what happens across the entire Fermi surface,
but very sensitive to the properties of the hot spots. To further
investigate the SC properties of the system, in Fig. \ref{fig_susceptibility}
we plot the temperature-dependent inverse pairing susceptibility $\chi_{\mathrm{pair}}^{-1}\left(T\right)$
for all band dispersions at their respective QCPs. We find that, for
a rather wide temperature range, the normalized susceptibilities $\chi_{\mathrm{pair}}^{-1}\left(T\right)/\chi_{\mathrm{pair}}^{-1}\left(3T_{c}\right)$
plotted as function of $T/T_{c}$ collapse onto a single curve, for
all values of $\delta/t$ and of $L$. As a result, it follows that
the pairing susceptibility must be of the form of Eq. (\ref{chi_pair}).
While the constant $A_{\mathrm{pair}}$, which determines the overall
amplitude of the SC fluctuations, depends weakly on $\delta/t$ (see
Fig. \ref{fig_susceptibility}), the function $f_{\mathrm{pair}}\left(T/T_{c}\right)$,
which determines the temperature dependence of the SC fluctuations,
is universal and independent on the band dispersion. Therefore, these
results imply that for a wide range of temperatures, the SC fluctuation
spectrum is determined by the same energy scale that determines $T_{c}$
\textendash{} which, according to the analysis in Fig. \ref{fig_TBKT},
is related to the hot-spots properties.

\section{Comparison with the hot-spots Eliashberg analytical approximation}

To gain a deeper understanding of the origin of our QMC results, we
analytically solve the spin-fermion model within the hot-spots Eliashberg
approximation introduced in previous works \cite{Chubukov03,Chubukov16,Kang16}.
Physically, the main assumptions of this approximation are that the
magnetic degrees of freedom are much slower than the electronic degrees
of freedom, and that the pairing instability arises only from the
hot spots (see Appendix A for technical details). Formally, the first
assumption can be justified if the number of electronic ``flavors''
is extended from $1$ to $N$, and $N$ is taken to be infinitely
large \textendash{} although recent works have raised important issues
on the general validity of a $1/N$ expansion \cite{SSLee09,Sachdev10,Senthil10}.

One of the main outcomes of the hot-spots Eliashberg approximation
is that the dynamics of the quantum magnetic fluctuations ceases to
be ballistic and, instead, becomes overdamped due to the decay of
spin fluctuations into electron-hole excitations \cite{Trebst16}.
The strength of this process is encoded in the Landau damping parameter
$\gamma\propto v_{F}^{2}\sin\theta_{\mathrm{hs}}/\lambda^{2}$, which
depends on the Fermi velocity at the hot spots $v_{F}$, on the interaction
parameter $\lambda$, and on the hot-spot angle $\sin\theta_{\mathrm{hs}}$.
The latter is nothing but a constraint on the phase space available
for the decay of the spin fluctuations into electron-hole pairs. This
property already suggests that the dependence of $T_{c}$ on $\sin\theta_{\mathrm{hs}}$
observed in the QMC results must be connected to the Landau damping.
Indeed, a full analysis reveals that, at the QCP, the only energy
scale in the hot-spots Eliashberg approximation is given by:

\begin{equation}
\Lambda_{\mathrm{QCP}}\propto\left(\frac{\lambda^{2}}{v_{F}}\right)^{2}\gamma\propto\lambda^{2}\sin\theta_{\mathrm{hs}}\label{Lambda_QCP1}
\end{equation}
which does not depend on the density of states or the Fermi velocity.
Consequently, the superconducting transition temperature at the QCP
can only depend on this energy scale \cite{Kang16,Chubukov16}, yielding
$T_{c}^{(\mathrm{hs})}=A_{c}^{(\mathrm{hs})}\lambda^{2}\sin\theta_{\mathrm{hs}}$,
in agreement with the QMC results. We use the superscript $(\mathrm{hs})$
to distinguish the calculated $T_{c}^{(\mathrm{hs})}$ from the numerically
obtained $T_{c}$. If we plug in the bare value of the interaction
parameter on the hot-spots Eliashberg approximation, we obtain $T_{c}^{(\mathrm{hs})}/t=0.14\sin\theta_{\mathrm{hs}}$,
which is very close to the linear fitting in Fig. \ref{fig_TBKT},
$T_{c}^{(\mathrm{hs})}/t=0.13\sin\theta_{\mathrm{hs}}$. However,
in comparing $T_{c}^{(\mathrm{hs})}$ with our QMC results, it is
important to recognize that the BKT physics is absent in the hot-spots
Eliashberg approximation. Of course, if the phase fluctuations responsible
for the suppression of $T_{c}^{(\mathrm{hs})}$ are only weakly sensitive
on the band structure parameters \cite{Schmalian05}, then the Eliashberg
transition temperature $T_{c}^{(\mathrm{hs})}$ and the BKT transition
temperature $T_{c}$ should be simply related by a constant $\alpha$,
$T_{c}=\alpha T_{c}^{(\mathrm{hs})}$. The fact that $T_{c}$ scales
linearly with $\sin\theta_{\mathrm{hs}}$ in our QMC simulations suggests
that this is indeed the case.

We can also compute the pairing susceptibility $\chi_{\mathrm{pair}}^{(\mathrm{hs})}\left(T\right)$
within the hot-spots Eliashberg approximation. At the QCP, we obtain
an expression of the form of Eq. (\ref{chi_pair}), with the universal
function $f_{\mathrm{pair}}^{(\mathrm{hs})}(T/T_{c})$ plotted together
with the collapsed QMC points in Fig. \ref{fig_susceptibility}. The
overall agreement between the two curves is evident and, surprisingly,
holds over a rather wide temperature range. This confirms our previous
conclusion that $f_{\mathrm{pair}}(x)$ arises from hot-spots properties.
The fact that the analytical function $f_{\mathrm{pair}}^{(\mathrm{hs})}(T/T_{c})$,
which is insensitive to BKT physics, captures well the behavior of
the QMC-derived function $f_{\mathrm{pair}}(T/T_{c})$, suggests that
vortex-antivortex fluctuations characteristic of the BKT transition
do not play a major role in our QMC simulations. Indeed, for all system
sizes studied, $\chi_{\mathrm{pair}}\left(T\right)$ does not show
any indication of an exponential temperature dependence near $T_{c}$.

\begin{figure}
\begin{centering}
\includegraphics[width=0.9\columnwidth]{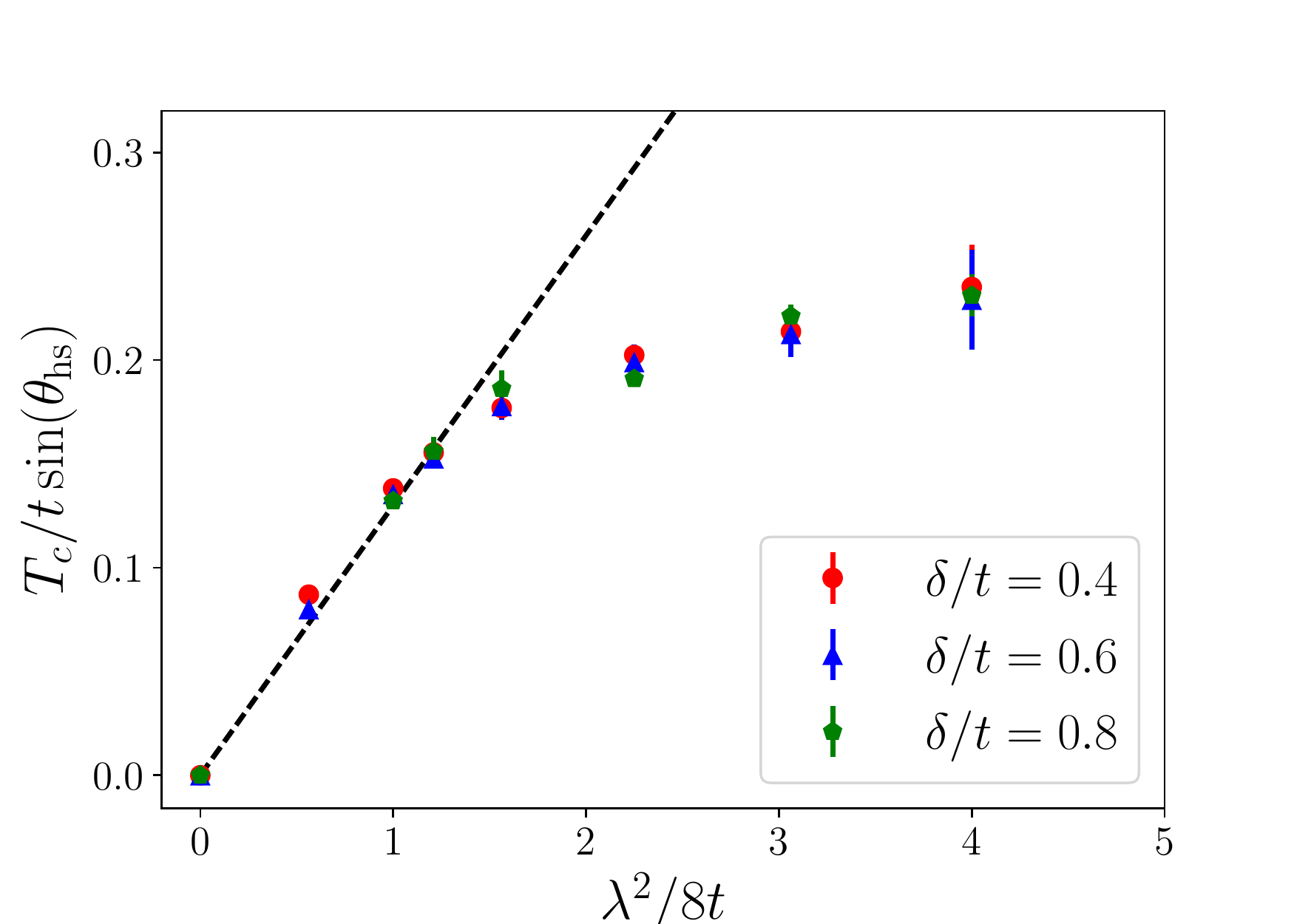} 
\par\end{centering}
\caption{Dependence of the superconducting transition temperature on the interaction
strength. For three values of the band dispersion parameter $\delta/t$,
we show the QMC results for $T_{c}$, in units of the hopping parameter
$t$ and normalized by the corresponding value of $\sin\theta_{\mathrm{hs}}$,
as function of the squared coupling constant $\lambda^{2}$ (in units
of $8t$) describing how strong the electrons interact with AFM fluctuations.
The system size is $L=12$. The dashed line, which denotes a $\lambda^{2}$
dependence, has the same slope as in Fig. \ref{fig_TBKT}b, and is
expected from the analytical hot-spots Eliashberg solution of the
spin-fermion model. The absence of the data point corresponding to
$\delta/t=0.8$ and $\lambda^{2}=4.5t$ is because $T_{c}$ did not
converge as function of the system size for these parameters. \label{fig_Yukawa}}
\end{figure}

An important prediction of the hot-spots Eliashberg approximation
is that $T_{c}$ increases not only with $\sin\theta_{\mathrm{hs}}$,
but also with $\lambda^{2}$. As a result, if the hot-spots Eliashberg
approximation is correct, $T_{c}$ would not be bounded and could
increase indefinitely as function of the interaction parameter $\lambda$.
To verify this property, we chose three band dispersion parameters
and obtained $T_{c}$ for several values of $\lambda$. As shown in
Fig. \ref{fig_Yukawa}, we find a reasonable scaling of $T_{c}/\sin\theta_{\mathrm{hs}}$
with $\lambda^{2}$ for moderately large values of the interaction
parameter, i.e. $\lambda^{2}$ of the order of the bandwidth $8t$.
The slope of this line is the same as that in Fig. \ref{fig_TBKT}b.
Note that for $\lambda=0$, we have a system of non-interacting electrons
with $\chi_{\mathrm{pair}}=2N_{f}\ln\left(\frac{\Lambda}{T}\right)$,
implying that $T_{c}=0$. More interestingly, for $\frac{\lambda^{2}}{8t}\gtrsim2$,
we start observing strong deviations from the $\lambda^{2}$ behavior,
signaling the failure of the hot-spots Eliashberg approximation. Furthermore,
in this regime, $T_{\mathrm{c}}$ increases very mildly and seems
to saturate.

To shed light on this behavior, we note that a key approximation of
the hot-spots Eliashberg approach is that the momentum associated
with the hot-spots typical energy scale \textendash{} also called
the hot-spots width, $\delta q_{\mathrm{hs}}\sim a^{-1}\sqrt{T_{c}/\gamma}$
\textendash{} is small compared to the Fermi momentum $q_{F}\sim1/a$.
However, because both $T_{c}$ and $\gamma^{-1}$ increase with $\lambda^{2}$,
the hot-spots width $\delta q_{\mathrm{hs}}$ also increases with
$\lambda^{2}$, and eventually becomes comparable to $q_{F}$ for
large enough values of $\lambda$. In this situation, the whole Fermi
surface becomes hot and effectively behaves as a ``large hot-spot.''
In this case, as shown in the supplementary material, the system still
has a single energy scale at the QCP, but instead of Eq. (\ref{Lambda_QCP1})
it is given by:

\begin{equation}
\tilde{\Lambda}_{\mathrm{QCP}}\propto p_{0}\left(\frac{\lambda^{2}}{v_{F}}\right)\gamma\propto p_{0}v_{F}\sin\theta_{\mathrm{hs}}\label{Lambda_QCP2}
\end{equation}
where $p_{0}$ is a momentum scale associated with the size of the
Fermi surface, and therefore is not a hot-spot property. Thus, in
this limit, $T_{c}^{(\mathrm{hs})}$ becomes independent of $\lambda$
and saturates. A similar behavior was found in Ref. \cite{Abanov08}
for the one-band spin-fermion model. Therefore, we can attribute the
near saturation of $T_{c}$ observed in our QMC results to a crossover
from pairing dominated by the hot spots to pairing dominated by the
entire Fermi surface. Naively, in the latter case, one would expect
$T_{c}$ to be more sensitive to the van Hove singularity. Interestingly,
our QMC results for $\frac{\lambda^{2}}{8t}=4$ do not reveal a sharp
enhancement near $\delta=t/4$ (see Appendix B). One possible reason
for this behavior is that the Fermi surface properties become less
important when interactions become too strong. While a detailed analysis
is beyond the scope of this paper \cite{Norman93}, future analytical
studies of the spin-fermion model near a van Hove singularity could
shed light on this behavior.

\section{Concluding remarks}

In summary, we showed that within the spin-fermion model the SC properties
near an AFM quantum critical point, including both the transition
temperature $T_{c}$ and the temperature-dependent pairing susceptibility
$\chi_{\mathrm{pair}}$, are dominated by the properties of the hot
spots, while being rather insensitive to the global properties of
the Fermi surface. More specifically, the functional dependences of
$T_{c}$ and $\chi_{\mathrm{pair}}$ inferred from our QMC results,
given by Eqs. (\ref{T_BKT}) and (\ref{chi_pair}), are very well
captured by an approximate analytical solution of the spin-fermion
model that focuses on the impact of the Landau damping on the pairing
interaction. In other words, the hot-spots Eliashberg approach provides
an excellent approximate solution to the spin-fermion model, which
presumably should hold also for systems with different types of band
dispersions beyond the rather artifical two-band case. It is surprising
that such an approximation works well even for moderately large values
of the interaction $\lambda^{2}$ between the AFM fluctuations and
the low-energy electronic states. However, our combined QMC-analytical
analysis also reveals that when $\lambda^{2}$ becomes larger than
the electronic bandwidth, the hot-spots approximation fails. Interestingly,
at this crossover from hot-spots dominated pairing to Fermi-surface
dominated pairing, $T_{c}$ seems to start saturating, signaling that
the maximum possible $T_{c}$ value for this model has been achieved.

Our results have important implications to the understanding of quantum
critical pairing in general. On the one hand, by establishing that
the properties of the hot spots govern the SC properties of the low-energy
spin-fermion model, it offers important insights into which of the
many system parameters should be changed to optimize $T_{c}$ in an
ideal system. For instance, it becomes clear that systems with nearly-nested
Fermi surfaces, where $\sin\theta_{\mathrm{hs}}$ is small, despite
having an abundance of low-energy magnetic fluctuations, have a much
smaller transition temperature than systems with non-nested Fermi
surfaces, where $\sin\theta_{\mathrm{hs}}$ is larger. Conversely,
our results establish robust and well-defined benchmarks that allow
one to assess whether the SC state obtained in other microscopic models
\textendash{} or even the superconducting state observed in actual
materials \textendash{} falls within the ``universality class''
of the low-energy spin-fermion model. Two such benchmarks, for instance,
are the linear dependence of $T_{c}$ on $\sin\theta_{\mathrm{hs}}$
and the saturation of $T_{c}$ for large interactions. Large-cluster
DMFT simulations of the Hubbard model \cite{Hubbard2,Hubbard3,Hubbard4}
may be able to test these benchmarks and elucidate whether the superconducting
properties of the Hubbard model are determined by hot-spots properties
or whether they depend on physics beyond the spin-fermion model. On
the experimental front, the most promising material candidates that
show signatures of AFM quantum criticality near optimal doping are
electron-doped cuprates and isovalent-doped pnictides. As for hole-doped
cuprates, although they do have a putative AFM quantum critical point,
they also display phenomena that have yet to be observed in QMC simulations
of the spin-fermion model, such as additional intertwined ordered
phases \cite{Fradkin15} and a transition from small to large Fermi
surface without an obvious accompanying order \cite{Taillefer16}.
One interesting possibility is to investigate how pressure affects
$T_{c}$ in these compounds, and correlate these changes with the
pressure-induced modifications of the hot-spots properties. 
\begin{acknowledgments}

We thank A. Chubukov, J. Kang, S. Kivelson, S. Lederer, and J. Schmalian for useful discussions. X.W. and R.M.F. were supported by the US Department of Energy, Office of Science, Basic Energy Sciences, under Award No. DE-SC0012336. R.M.F. and X.W. thank the Minnesota Supercomputing Institute (MSI) at the University of Minnesota, where part of the numerical computations was performed. R.M.F. also acknowledges partial support from the Research Corporation for Science Advancement via the Cottrell Scholar Award, and X.W. acknowledges support from the Doctoral Dissertation Fellowship offered by the University of Minnesota. E.B. was supported by the Israel Science Foundation under Grant No.~1291/12, by the US-Israel BSF under Grant No. 2014209, by a Marie Curie career reintegration grant, and by an Alon fellowship. R.M.F. and E.B. are grateful for the hospitality of the Aspen Center for Physics, where part of this work was developed. The Aspen Center for Physics is supported by National Science Foundation Grant No.~PHY-1066293.
\end{acknowledgments}

\appendix

\section{Spin-fermion model: hot-spots Eliashberg approximation}

\subsection{Calculation of $T_{c}$}

The hot-spots Eliashberg approximation consists basically of three
steps, see for instance Ref. \cite{Chubukov03,Chubukov16}: (i) the
bosonic self-energy $\Pi\left(\mathbf{q},\omega_{n}\right)$ is computed
within one loop; (ii) the normal and anomalous parts of the fermionic
self-energy $\Sigma\left(\mathbf{q},\omega_{n}\right)$ are solved
self-consistently within one loop, without vertex corrections; (iii)
the resulting gap equations are solved only at the hot spots. In this
approximation, the electronic band dispersions are linearized in the
vicinities of the hot spots, $\varepsilon_{i\mathbf{k}}\approx\mathbf{v}_{F,i}\cdot(\mathbf{k}-\mathbf{k}_{\mathrm{hs}})$.
For the specific band dispersions of our model, because the hot spots
are always along the diagonal $\left|k_{x}\right|=\left|k_{y}\right|$,
we have $\left|\mathbf{v}_{F,i}\right|=v_{F}$ for all hot spots,
with:

\begin{equation}
v_{F}=2t\sqrt{2\left[1-\left(\frac{\mu}{4t}\right)^{2}\right]\left[\left(\frac{\delta}{t}\right)^{2}+1\right]}
\end{equation}

Another quantity that also depends on the band dispersion parameter
$\delta$ is the angle between the Fermi velocities of a hot-spot
pair:

\begin{equation}
\sin\theta_{\mathrm{hs}}=\frac{2\left(\delta/t\right)}{1+\left(\delta/t\right)^{2}}
\end{equation}

Note that the angle is always defined such that $\sin\theta_{\mathrm{hs}}>0$.
After computing the one-loop bosonic self-energy, we find the renormalized
propagator:

\begin{equation}
\chi^{-1}(\mathbf{q},i\Omega_{n})=\tilde{r}+\mathbf{q}^{2}a^{2}+\frac{|\Omega_{n}|}{\gamma}\label{suscept}
\end{equation}
where $\tilde{r}=r-\Pi\left(0,0\right)$ is the renormalized mass
term, and the Landau damping coefficient is given by:

\begin{equation}
\gamma=\frac{\pi v_{F}^{2}\sin\theta_{\mathrm{hs}}}{\lambda^{2}N}
\end{equation}

Here, $\lambda$ is the Yukawa coupling constant, and $N$ is the
number of hot spots pairs, which in our model is $N=4$. To compute
the one-loop self-consistent self-energy, it is convenient to work
on Nambu space, defined by the spinors $\psi_{c,\mathbf{k}}^{\dagger}\equiv(c_{\mathbf{k}\uparrow}^{\dagger},c_{-\mathbf{k}\downarrow})$,
and $\psi_{2,\mathbf{k}}^{\dagger}\equiv(d_{\mathbf{k}+\mathbf{Q}\uparrow}^{\dagger},d_{-\mathbf{k}-\mathbf{Q}\downarrow})$.
The self-energy is then given by: 
\begin{equation}
\hat{\Sigma}_{c,k}=\frac{n_{b}\lambda^{2}}{\beta V}\sum_{p}\chi(k-p)\hat{G}_{d,p}\label{eq:self_energy}
\end{equation}
where $n_{b}=1,2,3$ for Ising, XY and Heisenberg spins, respectively.
Here, $\beta$ is the inverse temperature, $V=L^{2}$ is the volume
of the system, and $k=\left(\omega_{n},\mathbf{k}\right)$. Hereafter,
we will measure all momenta in units of the inverse lattice spacing
$1/a$. To proceed, we parametrize the fermionic self-energy as $\hat{\Sigma}_{i,k}=(1-Z_{i,k})i\omega_{n}\tau_{0}+\zeta_{i,\mathbf{k}}\tau_{3}+\phi_{i,k}\tau_{1}$,
where $\tau$ are Pauli matrices in Nambu space. The normal components
of the self-energy are thus expressed in terms of $Z_{i,k}$ and $\zeta_{i,k}$,
whereas the anomalous part, proportional to the superconducting gap,
is expressed in terms of $\phi_{i,k}$. From Dyson's equations, we
obtain the dressed Green's function: 
\begin{align}
\hat{G}_{i,k}=-\frac{Z_{i,k}i\omega_{n}+\varepsilon_{i,\mathbf{k}}\tau_{3}+\phi_{i,k}\tau_{1}}{Z_{i,k}^{2}\omega_{n}^{2}+\varepsilon_{i,\mathbf{k}}^{2}+\phi_{i,k}^{2}}\label{eq:Greens_function}
\end{align}
with renormalized $\varepsilon_{i,\mathbf{k}}\rightarrow\varepsilon_{i,\mathbf{k}}+\zeta_{i,\mathbf{k}}$.
Substitution back into Eq. (\ref{eq:self_energy}) and linearizing
in $\phi_{i}$, we find the self-consistent equations: 

\begin{align}
(1-Z_{1,k})i\omega_{n} & =-\frac{n_{b}\lambda^{2}}{\beta V}\sum_{\omega_{m},\mathbf{p}}\chi(\mathbf{k}-\mathbf{p},i\omega_{n}-i\omega_{m})\nonumber \\
 & \times\left(\frac{Z_{2,p}i\omega_{m}}{Z_{2,p}^{2}\omega_{m}^{2}+\varepsilon_{2,\mathbf{p}}^{2}}\right)\nonumber \\
\phi_{1,k} & =-\frac{n_{b}\lambda^{2}}{\beta V}\sum_{\omega_{m},\mathbf{p}}\chi(\mathbf{k}-\mathbf{p},i\omega_{n}-i\omega_{m})\nonumber \\
 & \times\left(\frac{\phi_{2,p}}{Z_{2,p}^{2}\omega_{m}^{2}+\varepsilon_{2,\mathbf{p}}^{2}}\right)\label{eq:eliash_eqn}
\end{align}

Analogous equations hold for $Z_{2,k}$ and $\phi_{2,k}$. Note that
in the Eliashberg approximation, the bosonic propagator $\chi$ is
not calculated self-consistently, i.e. the bosonic self-energy is
computed using the non-interacting Green's functions \cite{Chubukov03}.

To proceed, we solve these equations only at the hot spots, and therefore
ignore the momentum dependence of the quasi-particle weight $Z$ and
of the gap $\phi$. Within the Eliashberg approximation, we only need
to consider the variation of the bosonic propagator with respect to
the momentum parallel to the Fermi surface, $\chi(\mathbf{q},i\Omega_{n})\approx\chi(q_{\parallel},i\Omega_{n})$.
These key aspects of the hot-spots Eliashberg approximation highlight
the fact that the bosonic degrees of freedom are much slower than
the fermionic ones. Using these approximations, one can then perform
the integration over momentum in the previous expressions by changing
coordinates to $\left(p_{\parallel},p_{\perp}\right)$, i.e. momenta
parallel and perpendicular to the Fermi surface near the hot spots.
As a result, $\varepsilon_{i,\mathbf{p}}=v_{F}p_{\perp}$, and one
obtains: 
\begin{align}
Z(\omega_{n}) & =1+\frac{n_{b}\lambda^{2}T}{4v_{F}}\sum_{\omega_{m}}V_{\mathrm{pair}}\left(\omega_{n}-\omega_{m}\right)\frac{\mathrm{sign}(\omega_{m})}{\omega_{n}}\label{eq:simplified_eliash_eqn}\\
\phi(\omega_{n}) & =\frac{n_{b}\lambda^{2}T}{4v_{F}}\sum_{\omega_{m}}V_{\mathrm{pair}}\left(\omega_{n}-\omega_{m}\right)\frac{\phi(\omega_{m})}{Z(\omega_{m})|\omega_{m}|}
\end{align}

To write these expressions, we note that $Z_{1}=Z_{2}$, since the
Fermi velocities are the same at both points of the hot-spot pair,
and $\phi_{1}=-\phi_{2}$ is the only possible solution to the gap
equations. The pairing interaction is given by:

\begin{equation}
V_{\mathrm{pair}}\left(\Omega_{n}\right)=\int_{-\frac{p_{0}}{2}}^{\frac{p_{0}}{2}}\frac{\mathrm{d}p_{\parallel}}{\pi}\frac{1}{p_{\parallel}^{2}+\tilde{r}+|\Omega_{n}|/\gamma}
\end{equation}
where $p_{0}\sim\mathcal{O}(1)$ is an upper momentum cutoff related
to the size of the Fermi surface in the Brillouin zone. This momentum
scale is to be compared to the typical ``momentum width'' of the
hot spots, $\delta p_{\mathrm{hs}}=\sqrt{2\pi T_{c}/\gamma}$, determined
by comparing the frequency and momentum dependent terms in Eq. (\ref{suscept})
for the energy scale $\Omega_{n}=2\pi T_{c}$. In the hot-spots Eliashberg
approximation, $p_{0}\gg\delta p_{\mathrm{hs}}$, and we can replace
$p_{0}\rightarrow\infty$ in the previous expression, yielding:

\begin{equation}
V_{\mathrm{pair}}\left(\Omega_{n}\right)=\sqrt{\frac{1}{\tilde{r}+|\Omega_{n}|/\gamma}}\label{Vpair}
\end{equation}

Therefore, the Eliashberg equations become: 
\begin{align}
Z(\omega_{n}) & =1+\frac{1}{2\pi}\sqrt{\frac{\Lambda_{\mathrm{QCP}}}{T}}\sum_{\omega_{m}}\frac{1}{\sqrt{|n-m|+\frac{\tilde{r}\gamma}{2\pi T}}}\,\frac{\mathrm{sign}(\omega_{m})}{n+\frac{1}{2}}\label{eq:eliash_eqnv1}\\
\phi(\omega_{n}) & =\frac{1}{2\pi}\sqrt{\frac{\Lambda_{\mathrm{QCP}}}{T}}\sum_{\omega_{m}}\frac{1}{\sqrt{|n-m|+\frac{\tilde{r}\gamma}{2\pi T}}}\,\frac{\phi(\omega_{m})}{Z(\omega_{m})|m+\frac{1}{2}|}
\end{align}
where we introduced the energy scale:

\begin{equation}
\Lambda_{\mathrm{QCP}}\equiv\left(\frac{n_{b}\lambda^{2}}{4v_{F}}\right)^{2}\frac{\gamma}{2\pi}=\frac{n_{b}^{2}\lambda^{2}\sin\theta_{\mathrm{hs}}}{32N}\label{Lambda_QCP}
\end{equation}

The key point is that at the QCP, $\tilde{r}=0$, and the only energy
scale in the problem is given by $\Lambda_{\mathrm{QCP}}$ (a similar
behavior is found slightly away from the QCP, as long as $\tilde{r}\ll2\pi T_{c}/\gamma$).
Therefore, the superconducting transition temperature at the QCP is
set by the only energy scale in the problem, i.e. $T_{c}=\alpha\Lambda_{\mathrm{QCP}}$,
where $\alpha$ is a number (no cutoff is necessary, in contrast to
the BCS case). According to our numerical solution of the Eliashberg
equations, we find $\alpha\approx0.56$, in agreement with previous
calculations \cite{Chubukov16,Kang16}. Note that, as pointed out
in Ref. \cite{Chubukov16}, when $\tilde{r}=0$, the term $m=n$ in
the sum that appears in the determination of $Z\left(\omega_{n}\right)$
is exactly canceled by the term $m=n$ in the sum that appears in
the determination of $\phi\left(\omega_{n}\right)$. This is easily
seen by defining the pairing gap $\Delta\equiv\phi/Z$, and separating
out the $m=n$ term from Eqs. (\ref{eq:eliash_eqnv1}): \begin{widetext}
\begin{align}
\left[Z(\omega_{n})-\frac{1}{2\pi}\sqrt{\frac{\Lambda_{\mathrm{QCP}}}{T}}\sqrt{\frac{2\pi T}{\tilde{r}\gamma}}\frac{1}{|n+\frac{1}{2}|}\right] & =1+\frac{1}{2\pi}\sqrt{\frac{\Lambda_{\mathrm{QCP}}}{T}}\sum_{\omega_{m}\neq\omega_{n}}\frac{1}{\sqrt{|n-m|+\frac{\tilde{r}\gamma}{2\pi T}}}\,\frac{\mathrm{sign}(\omega_{m})}{n+\frac{1}{2}}\label{eliash_eqnv2}\\
\Delta(\omega_{n})\left[Z(\omega_{n})-\frac{1}{2\pi}\sqrt{\frac{\Lambda_{\mathrm{QCP}}}{T}}\sqrt{\frac{2\pi T}{\tilde{r}\gamma}}\frac{1}{|n+\frac{1}{2}|}\right] & =\frac{1}{2\pi}\sqrt{\frac{\Lambda_{\mathrm{QCP}}}{T}}\sum_{\omega_{m}\neq\omega_{n}}\frac{1}{\sqrt{|n-m|+\frac{\tilde{r}\gamma}{2\pi T}}}\,\frac{\Delta(\omega_{m})}{|m+\frac{1}{2}|}
\end{align}
\end{widetext}Therefore the $m=n$ term does not enter into the linearized
gap equation, and that there is a finite superconducting transition
temperature in the limit $r\rightarrow0$. It is important to note
that $\lambda^{2}/N$ does not necessarily have the same bare value
that enters the Hamiltonian in the QMC simulations, since magnetic
fluctuations are known to effectively renormalize the interactions.
If nevertheless we use the bare values of $\lambda$ and $N$ to estimate
$T_{c}$, i.e. $\lambda^{2}=8t$ and $N=4$, we would get $T_{c}/t\approx0.14\,\sin\theta_{\mathrm{hs}}$,
which is about $10\%$ larger than the BKT superconducting transition
temperature obtained from the QMC simulations.

It is also instructive to consider the opposite limit in which the
entire Fermi surface becomes hot, i.e. \textbf{$p_{0}\ll\delta p_{\mathrm{hs}}$}.
This is certainly the case when $2\pi T_{c}\gg\gamma$; since $T_{c},\gamma^{-1}\propto\lambda^{2}$,
this means that this limit is achieved for large values of the Yukawa
coupling. In this case, the pairing interaction becomes: 
\begin{equation}
V_{\mathrm{pair}}\left(\Omega_{n}\right)=\frac{p_{0}/\pi}{\tilde{r}+|\Omega_{n}|/\gamma}
\end{equation}

As a result, at the QCP, $\tilde{r}=0$, there is still only one energy
scale in the Eliashberg equations, now set by: 
\begin{equation}
\tilde{\Lambda}_{\mathrm{QCP}}\equiv\frac{p_{0}}{\pi}\left(\frac{n_{b}\lambda^{2}}{4v_{F}}\right)\frac{\gamma}{2\pi}=\frac{n_{b}p_{0}v_{F}\sin\theta_{\mathrm{hs}}}{8\pi N}
\end{equation}
The Eliashberg equations become: 
\begin{align}
Z(\omega_{n}) & =1+\frac{1}{2\pi}\left(\frac{\tilde{\Lambda}_{\mathrm{QCP}}}{T}\right)\sum_{\omega_{m}}\frac{1}{|n-m|+\frac{\tilde{r}\gamma}{2\pi T}}\,\frac{\mathrm{sign}(\omega_{m})}{n+\frac{1}{2}}\label{eq:eliash_eqnv2}\\
\phi(\omega_{n}) & =\frac{1}{2\pi}\left(\frac{\tilde{\Lambda}_{\mathrm{QCP}}}{T}\right)\sum_{\omega_{m}}\frac{1}{|n-m|+\frac{\tilde{r}\gamma}{2\pi T}}\,\frac{\phi(\omega_{m})}{Z(\omega_{m})|m+\frac{1}{2}|}
\end{align}

Therefore, at $\tilde{r}=0$, $T_{c}=\tilde{\alpha}\tilde{\Lambda}_{\mathrm{QCP}}$
becomes independent of the Yukawa coupling, and may depend on additional
properties of the Fermi surface, as indicated by the presence of the
momentum scale $p_{0}$ in $\tilde{\Lambda}_{\mathrm{QCP}}$. Note
that due to similar arguments described in Eq.~\ref{eliash_eqnv2},
$n=m$ term does not appear in the linearized gap equation.

\subsection{Calculation of the pairing susceptibility}

To compute the static pairing susceptibility in the sign-changing
gap channel, we first introduce in the Hamiltonian the pairing field
$\Delta$: 
\begin{equation}
\delta H=-2\Delta\sum_{\mathbf{k}}\left(c_{\mathbf{k}\uparrow}c_{-\mathbf{k}\downarrow}-d_{\mathbf{k}\uparrow}d_{-\mathbf{k}\downarrow}+h.c.\right)
\end{equation}

Here the factor of $2$ is included so that the definition of the
pairing vertex is consistent with that used in the QMC simulations.
In Dyson's equation, this term can be incorporated in the self-energy,
$\hat{\Sigma}_{i}\rightarrow\hat{\Sigma}_{i}-2\Delta\tau_{1}$. Repeating
the same steps as above, the only modification is in the gap equation:
\begin{equation}
\phi(\omega_{n})=\frac{n_{b}\lambda^{2}T}{4v_{F}}\sum_{\omega_{m}}V_{\mathrm{pair}}\left(\omega_{n}-\omega_{m}\right)\frac{\phi(\omega_{m})}{Z(\omega_{m})|\omega_{m}|}+2\Delta
\end{equation}

We considered the linearized equation because we are interested only
in the susceptibility of the disordered state, where $\phi=0$. Defining
$\eta(\omega_{n})\equiv\partial\phi(\omega_{n})/\partial\Delta$,
we obtain a self-consistent equation for $\eta\left(\omega_{n}\right)$:
\begin{align}
\eta(\omega_{n}) & =\frac{n_{b}\lambda^{2}T}{4v_{F}}\sum_{\omega_{m}}V_{\mathrm{pair}}\left(\omega_{n}-\omega_{m}\right)\frac{\eta(\omega_{m})}{Z(\omega_{m})|\omega_{m}|}+2\label{eq:pair_sus_calculation}\\
\eta(\omega_{n}) & =\frac{1}{2\pi}\sqrt{\frac{\Lambda_{\mathrm{QCP}}}{T}}\sum_{\omega_{m}}\frac{1}{\sqrt{|n-m|}}\,\frac{\eta(\omega_{m})}{Z(\omega_{m})|m+\frac{1}{2}|}+2
\end{align}

Now, the static pairing susceptibility is given by:

\begin{equation}
\chi_{\mathrm{pair}}\equiv\chi(\mathbf{q}\rightarrow0,i\Omega_{n}\rightarrow0)=\partial_{\Delta}\sum_{k}2\langle c_{k\uparrow}c_{-k\downarrow}-d_{k\uparrow}d_{-k\downarrow}\rangle
\end{equation}
where $k=\left(\omega_{n},\mathbf{k}\right)$ and $\sum_{k}=T\sum_{n}\int\frac{d^{2}k}{\left(2\pi\right)^{2}}$.
Because the mean value is precisely minus the anomalous part of the
Green's function, $\phi_{i,k}$, given by Eq. (\ref{eq:simplified_eliash_eqn}),
we obtain:

\begin{equation}
\chi_{\mathrm{pair}}=4T\sum_{\omega_{n}}\int\frac{d^{2}k}{\left(2\pi\right)^{2}}\frac{\eta(\omega_{n})}{Z(\omega_{n})^{2}\omega_{n}^{2}+\varepsilon_{\mathbf{k}}^{2}}
\end{equation}

Note that, for $\lambda=0$ (non-interacting electrons), we have $Z\left(\omega_{n}\right)=1,\ \eta\left(\omega_{n}\right)=2$,
and the equation above reduces to the well-known BCS expression 
\begin{align}
\chi_{\mathrm{pair}} & =8T\sum_{\omega_{n},\mathbf{k}}\frac{1}{\omega_{n}^{2}+\varepsilon_{\mathbf{k}}^{2}}\label{eq:non_int_pair_sus}\\
 & =4\sum_{i,k}G_{i}^{(0)}\left(k\right)G_{i}^{(0)}\left(-k\right)\nonumber \\
 & =2N_{f}\ln\frac{\Lambda}{T}\nonumber 
\end{align}
where $N_{f}=4\int\frac{\mathrm{d^{2}k}}{(2\pi)^{2}}\delta(\varepsilon_{\mathbf{k}})$
is the total density of states at the Fermi level. The factor of $4$
arises due to band and spin degeneracies.

For $\lambda\neq0$, it is convenient once again to integrate along
directions parallel and perpendicular to the Fermi surface, yielding:

\begin{equation}
\chi_{\mathrm{pair}}=\frac{p_{0}}{2\pi^{2}v_{F}}\sum_{\omega_{n}}\frac{\eta(\omega_{n})}{Z(\omega_{n})\left|n+\frac{1}{2}\right|}
\end{equation}
where $p_{0}$ is the same quantity as defined in the previous section.
Because the equations for $\eta\left(\omega_{n}\right)$ and $Z\left(\omega_{n}\right)$,
Eqs. (\ref{eq:pair_sus_calculation}) and (\ref{eq:eliash_eqnv1}),
depend only on $T/\Lambda_{\mathrm{QCP}}\propto T/T_{c}$, it follows
that the susceptibility is of the form $\chi_{\mathrm{pair}}\left(T\right)=A_{\mathrm{pair}}f_{\mathrm{pair}}\left(\frac{T}{T_{c}}\right)$,
where $A_{\mathrm{pair}}$ depends on the Fermi surface properties
(as signaled by $p_{0}$ above), but $f_{\mathrm{pair}}\left(\frac{T}{T_{c}}\right)$
is a universal function.

In computing $\chi_{\mathrm{pair}}$ numerically, it is important
to keep in mind that as higher temperatures are considered, the effect
of the bandwidth becomes more important, as the bandwidth $8t$ provides
a natural energy cutoff for the Matsubara sum. Note that this is not
an issue for the computation of $T_{c}$, since $T_{c}\ll8t$ always.
Because $8t$ is a hard cutoff in real frequency space, to capture
its effects in Matsubara frequency space, we follow Ref. \cite{Kang16}
and introduce a soft cutoff:

\begin{equation}
\Upsilon(\omega_{n})=\frac{1}{\exp\left[(\omega_{n}-8t)/\omega_{0}\right]+1}
\end{equation}

This function appears not only in the Matsubara sum present in $\chi_{\mathrm{pair}}$,
but also in the self-consistent equation for $\zeta\left(\omega_{n}\right)$
via:

\begin{align}
\eta(\omega_{n}) & =2\Upsilon(\omega_{n})+\frac{1}{2\pi}\sqrt{\frac{\Lambda_{\mathrm{QCP}}}{T}}\nonumber \\
 & \times\left[\sum_{\omega_{m}}\frac{\Upsilon(\omega_{n})\Upsilon(\omega_{m})}{\sqrt{|n-m|}}\,\frac{\eta(\omega_{m})}{Z(\omega_{m})|m+\frac{1}{2}|}\right]
\end{align}

For the plot in Fig. 4b of the main text, we used $\omega_{0}=1.6t$.
Changing this parameter slightly does not affect the main properties
of $\chi_{\mathrm{pair}}$.

\section{Determinant Quantum Monte Carlo }

The technical details of the implementation of the determinant Quantum
Monte Carlo (QMC) for the two-band spin-fermion model with XY spins
are the same as those extensively presented in Ref. \cite{Schattner16},
co-authored by two of us. As explained in the main text, in this work
our goal is to establish the band structure parameters that determine
$T_{c}$ and $\chi_{\mathrm{pair}}$. Our procedure is the following:
for a given set of parameters, we first determine the approximate
location of the AFM quantum critical point $r_{c}$ and then determine
$T_{c}$ from the condition that the superfluid density $\rho_{s}$
reaches the BKT value $2T_{c}/\pi$. The static pairing susceptibility
is computed directly\emph{.} In this supplementary section, we provide
more details of how these three quantities are determined for a given
set of parameters $\left(\delta,\ L,\:\lambda^{2}\right)$, characterized
by the band parameter $\delta/t=0.2,\,0.3,\,0.4,\,0.5,\,0.6,\,0.7,\,0.8$,
the system size $L=8,\,10,\,12,\,14$, and the squared coupling constant
$\lambda^{2}/t=8$.

\subsection{Antiferromagnetic quantum critical point (AFM-QCP)}

The AFM-QCP is reached by tuning the bare mass term of the magnetic
propagator to $r=r_{c}$, see Eq. (6) of the main text. Determining
the precise location of the QCP is a very difficult task, not only
due to the BKT character of the AFM transition at finite temperatures
(since we are dealing with XY spins), but also because once superconductivity
sets in, it competes with AFM order and shifts the location of the
QCP from $r_{c}$ to $\bar{r}_{c}<r_{c}$. This last behavior was
indeed observed in the previous QMC studies of Ref. \cite{Schattner16}.
However, for our purposes, it is not necessary to precisely determine
the position $r_{c}$ of the QCP. As explained in the previous section,
the onset of superconductivity within the hot-spots Eliashberg approximation
of the spin-fermion model depends on two parameters, $\Lambda_{\mathrm{QCP}}\propto\lambda^{2}\sin\theta_{\mathrm{hs}}$,
and the renormalized mass of the magnetic propagator, $\tilde{r}$,
see for instance Eqs. (\ref{Vpair}) and (\ref{eq:eliash_eqnv1}).
Thus, as long as $\tilde{r}\ll2\pi T_{c}/\gamma$, the superconducting
properties of the system are effectively the same as those at the
QCP. Therefore, to probe quantum critical pairing, we search for a
value of $r$ sufficiently close to $r_{c}$ such that $\tilde{r}$
is very small, but non-zero, since we must ensure that the system
is not in the AFM ordered phase.

For this purpose, we first define the uniform magnetization: 
\begin{equation}
\bar{\mathbf{M}}\equiv\frac{1}{\beta L^{2}}\sum_{\mathbf{r}}\int d\tau\,\mathbf{M}(\mathbf{r},\tau)
\end{equation}

To obtain a good estimate of $r_{c}$, we extract from the QMC simulations
both the Binder cumulant, 
\begin{equation}
\mathcal{B}=1-\frac{\langle\left(\bar{\mathbf{M}}^{2}\right)^{2}\rangle}{2\langle\bar{\mathbf{M}}^{2}\rangle^{2}}
\end{equation}
and the static spin susceptibility, 
\begin{equation}
\chi_{M}\equiv\frac{1}{\beta L^{2}}\langle\sum_{\mathbf{r},\tau}\sum_{\mathbf{r}^{\prime},\tau^{\prime}}\mathbf{M}(\mathbf{r},\tau)\cdot\mathbf{M}(\mathbf{r}^{\prime},\tau^{\prime})\rangle=\beta L^{2}\langle\bar{\mathbf{M}}^{2}\rangle
\end{equation}

\begin{figure}
\includegraphics[width=1\columnwidth]{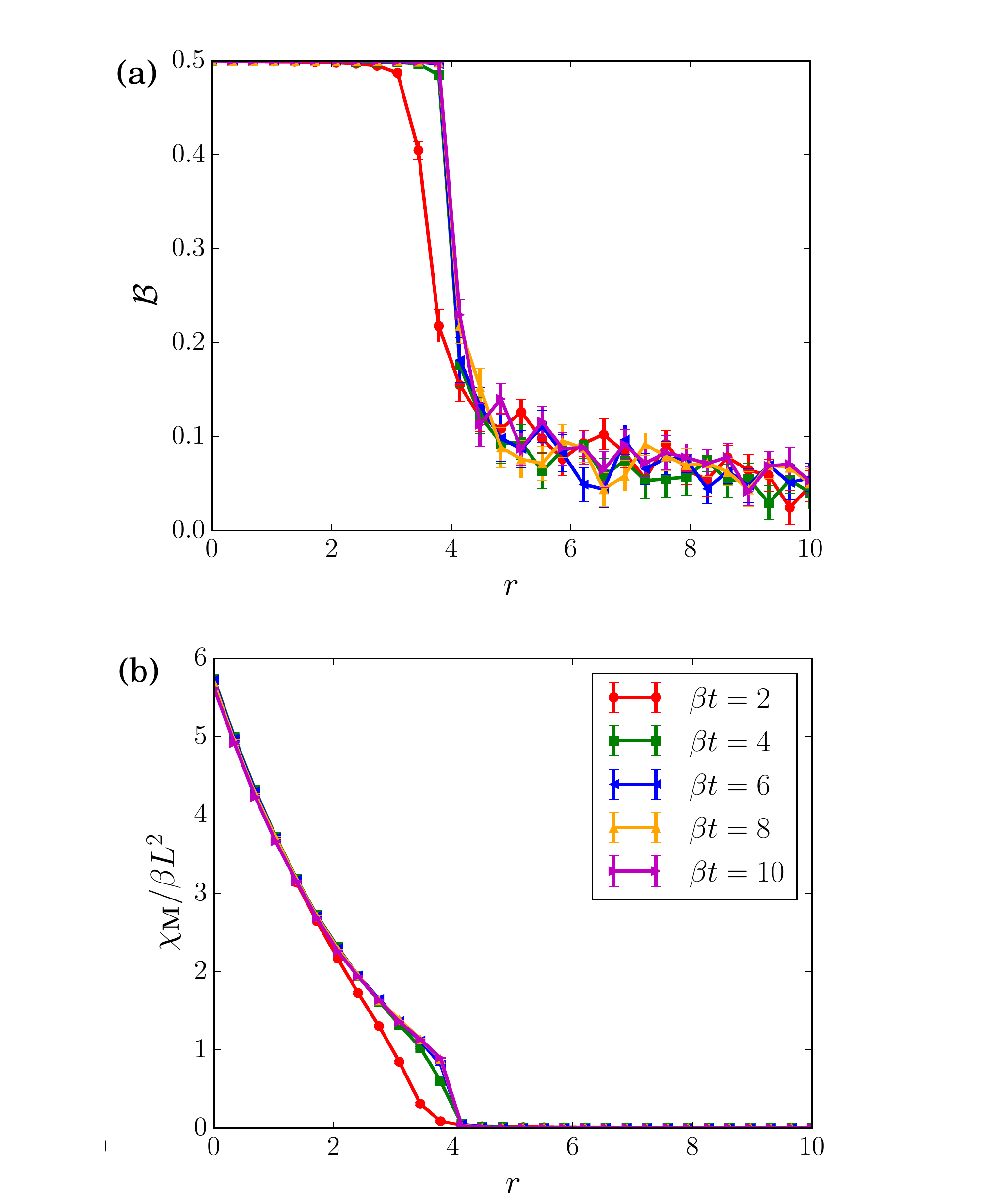}\caption{\label{fig:binder_spin}Binder cumulant $\mathcal{B}$ (a) and
static spin susceptibility $\chi_{M}$ (b) as a function of $r$
for various inverse temperatures. The set of parameters used here
is $\left(\delta/t,\ L,\:\lambda^{2}/t\right)=\left(0.6,\,12,\,8\right)$.
The inverse temperature $\beta$ is in units of $1/t$.}
\end{figure}

Here, $\langle\cdots\rangle$ denotes thermal averaging. For XY spins
deep in the ordered phase, $\mathcal{B}=\frac{1}{2}$, whereas $\mathcal{B}=0$
deep in the disordered phase. Similarly, in the ordered phase, $\chi_{M}$
scales with $\beta L^{2-\eta}$, where $\eta$ changes continuously
as function of $r$ and $T$, approaching $\eta=0$ deep in the ordered
phase. Therefore, at any finite temperature, a rough estimate for
the AFM transition is given by the value of $r$ in which $\chi_{M}/\left(\beta L^{2}\right)$
shows a kink and $\mathcal{B}$ changes sharply from $0$ to $1/2$.
In Fig. \ref{fig:binder_spin}, we show the behavior of these two
quantities, plotted as function of $r$ for different fixed temperatures,
for the set of parameters $\left(\delta/t,\ L,\:\lambda^{2}/t\right)=\left(0.6,\,12,\,8\right)$.
On the scale shown in this figure, both $\mathcal{B}$ and $\chi_{M}/\left(\beta L^{2}\right)$
are nearly temperature independent at low temperatures (but still
above $T_{c}$), therefore providing an estimate for $r_{c}$.

\begin{figure}
\begin{centering}
\includegraphics[width=0.5\columnwidth]{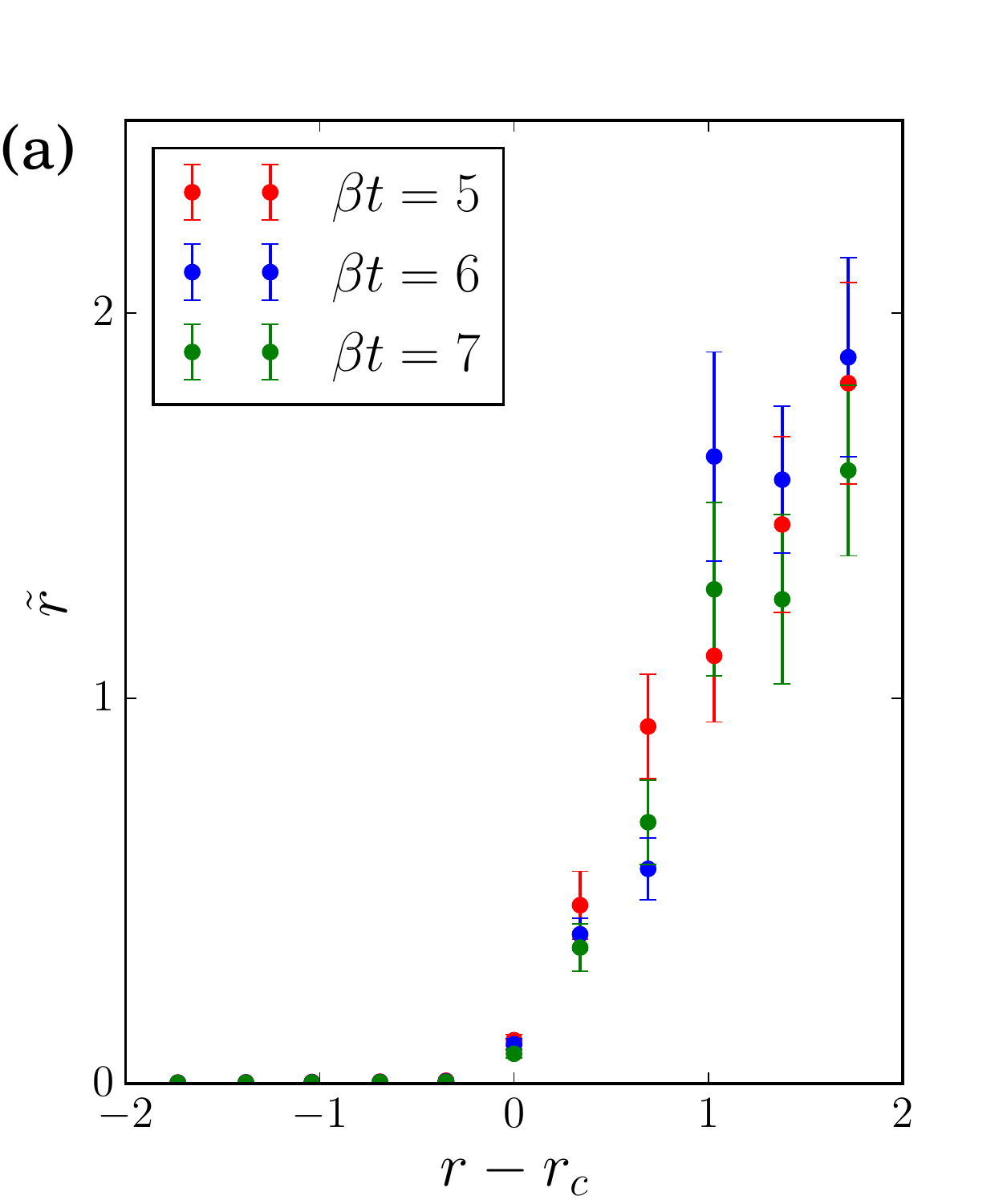}
\par\end{centering}
\bigskip{}

\centering{}\includegraphics[width=1\columnwidth]{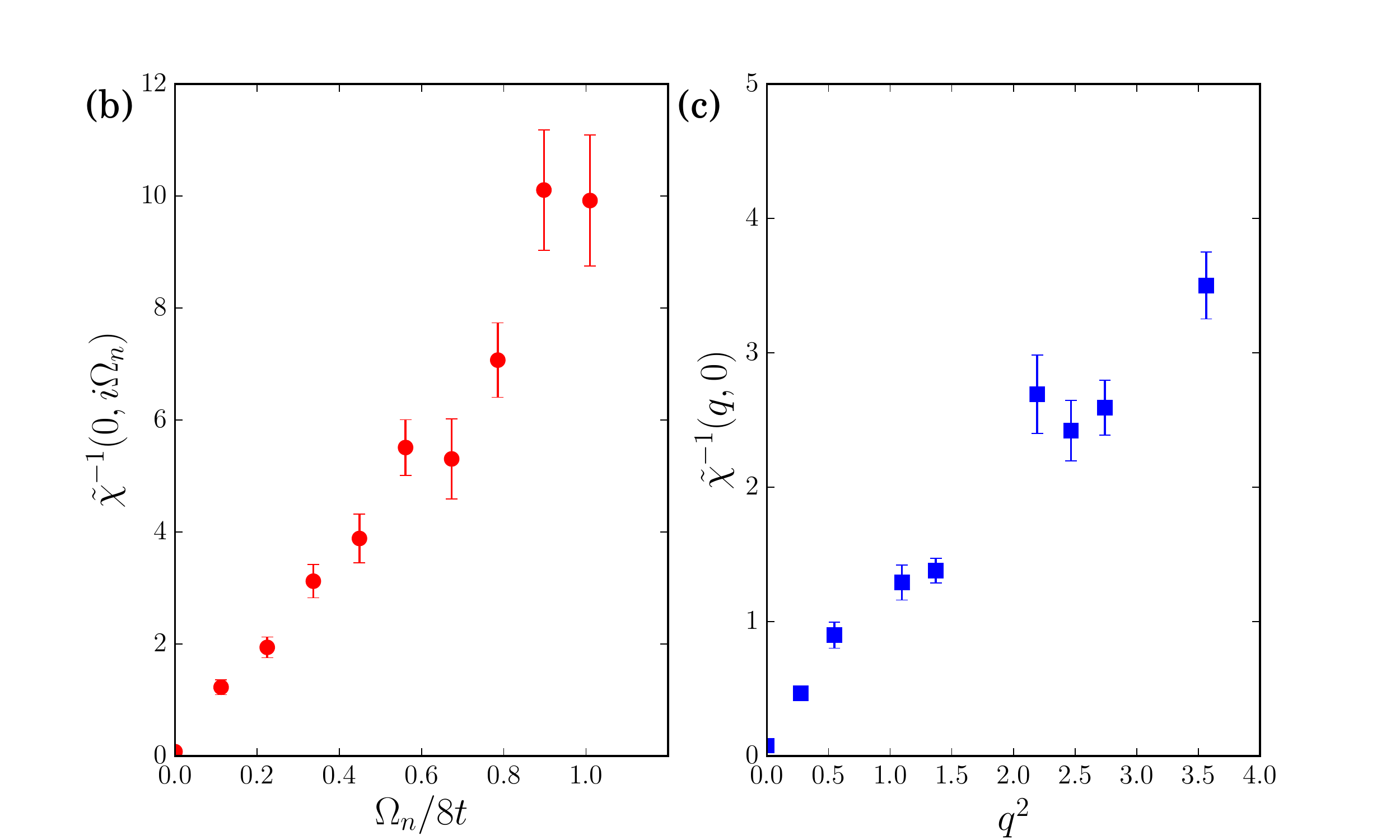}\caption{\label{fig:dprop}Panel (a) shows the renormalized mass term of the
magnetic propagator, $\tilde{r}$, as function of $r_{c}$. The set
of parameters used here is $\left(\delta/t,\ L,\:\lambda^{2}/t\right)=\left(0.6,\,12,\,8\right)$.
The inverse renormalized magnetic propagator $\tilde{\chi}^{-1}(\mathbf{q},i\Omega_{n})$
at $r=r_{c}$ is plotted as function of $\Omega_{n}$ for $\mathbf{q}=0$
(b) and as function of $q$ for $\Omega_{n}=0$ (c). In (b) and (c),
the inverse temperature is $\beta=7/t$.}
\end{figure}

Next, to improve our estimate of $r_{c}$, we compute the $r$ dependence
of the mass of the bosonic propagator at low temperatures, $\tilde{r}\equiv\chi_{M}^{-1}\left(\mathbf{q}=0,i\Omega_{n}=0\right)$,
as shown in Fig. \ref{fig:dprop}(a). The estimated $r_{c}$ corresponds
to the $r$ value that has the smallest $\tilde{r}$, before however
it reaches zero, since we want to study the system in the non-magnetically
ordered state.

In the same figure we also present the frequency and momentum dependencies
of $\chi_{M}^{-1}\left(\mathbf{q},i\Omega_{n}\right)$. In agreement
to a recent study by some of us \cite{Trebst16}, $\chi_{M}^{-1}(\mathbf{q}=0,i\Omega_{n})$
shows a rather linear dependence on the Matsubara frequency, indicating
the presence of Landau damping, which in turn plays a key role in
the hot-spots Eliashberg approximation, see Eq. (\ref{suscept}).
Similarly, $\chi_{M}^{-1}(\mathbf{q},i\Omega_{n}=0)$ is consistent
with a $q^{2}$ behavior for small momentum.

\subsection{Pairing susceptibility and superfluid density}

\begin{figure}
\includegraphics[width=1\columnwidth]{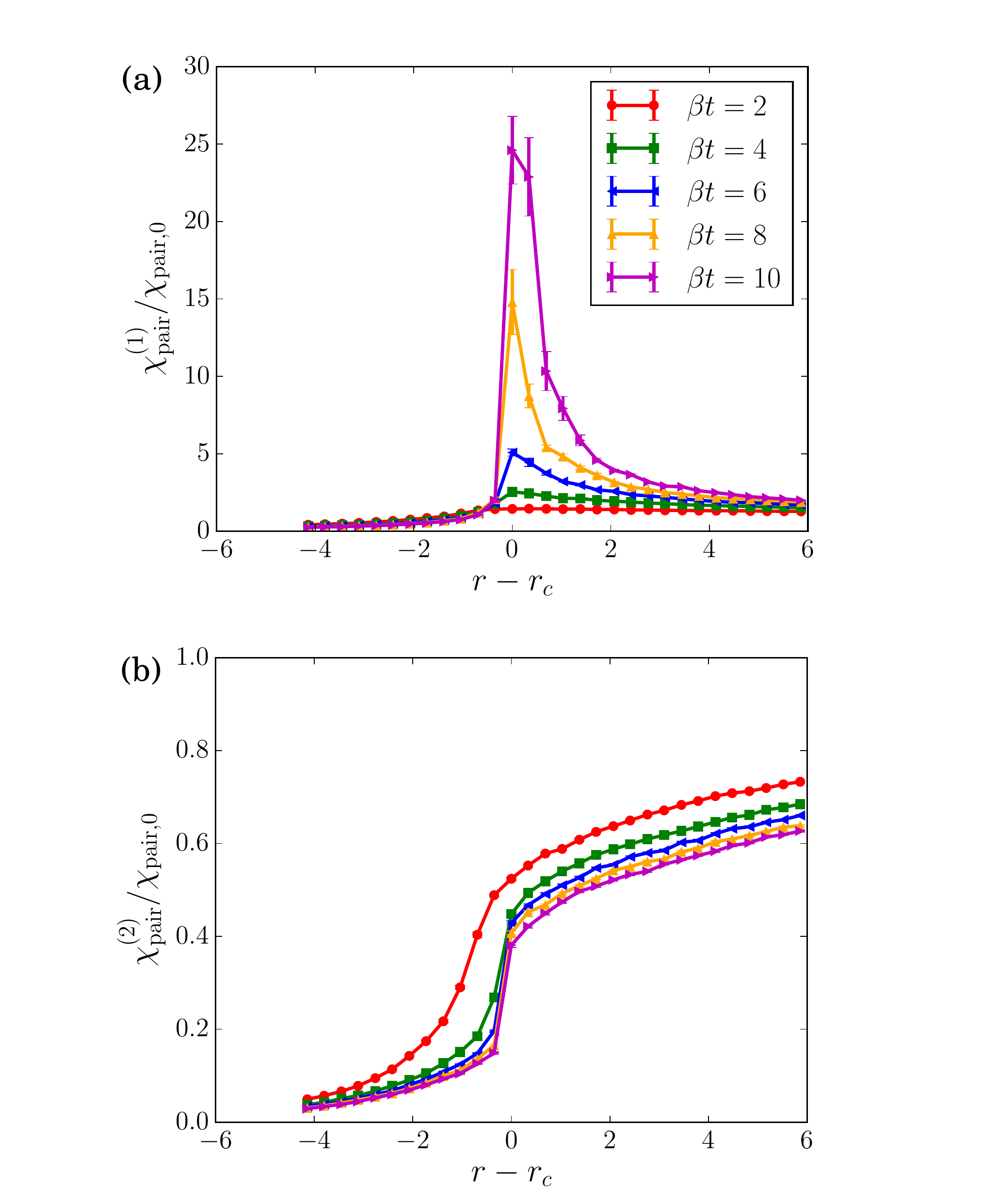}\caption{\label{fig:fermion_sector}Static pairing susceptibility $\chi_{\mathrm{pair}}^{(a)}$
in the sign-changing gap channel ($a=1$, panel a) and in the in the
sign-preserving gap channel ($a=2$, panel b) as function of the distance
to the QCP at $r=r_{c}$. The inverse temperature $\beta$ is in units
of $1/t$ and the susceptibilities are normalized by the non-interacting
susceptibility $\chi_{\mathrm{pair,0}}$ obtained by setting $\lambda=0$.
The set of parameters used here is $\left(\delta/t,\ L,\:\lambda^{2}/t\right)=\left(0.6,\,12,\,8\right)$. }
\end{figure}

The static pairing susceptibility is defined as: 
\begin{equation}
\chi_{\text{pair}}^{(a)}\equiv\frac{1}{\beta L^{2}}\sum_{\mathbf{r},\mathbf{r}^{\prime}}\int_{\tau,\tau'}\langle\Gamma_{a}(\mathbf{r},\tau)\Gamma_{a}^{\dagger}(\mathbf{r}^{\prime},\tau^{\prime})\rangle
\end{equation}
where

\begin{equation}
\Gamma_{a}(\mathbf{r},\tau)\equiv i\sigma_{\alpha\beta}^{y}\left[c_{\alpha}(\mathbf{r},\tau)c_{\beta}(\mathbf{r},\tau)+(-1)^{a}d_{\alpha}(\mathbf{r},\tau)d_{\beta}(\mathbf{r},\tau)\right]
\end{equation}
is the pairing field associated with the sign-changing gap function
$(a=1)$ or to the sign-preserving gap function $(a=2)$. $\sigma_{y}$
is the Pauli matrix in spin space. In Fig. \ref{fig:fermion_sector},
we plot both pairing susceptibilities, in units of the non-interacting
susceptibility $\chi_{\mathrm{pair,0}}$, as function of $r$ and
as function of temperature for the set of parameters $\left(\delta/t,\ L,\:\lambda^{2}/t\right)=\left(0.6,\,12,\,8\right)$.
Compared with Fig. \ref{fig:binder_spin}, it is clear that while
$\chi_{\mathrm{pair}}^{(1)}/\chi_{\mathrm{pair,0}}$ is strongly peaked
at $r=r_{c}$, $\chi_{\mathrm{pair}}^{(2)}/\chi_{\mathrm{pair,0}}$
is always smaller than $1$, implying that there is no enhancement
in the sign-preserving channel.

\begin{figure}
\includegraphics[width=0.9\columnwidth]{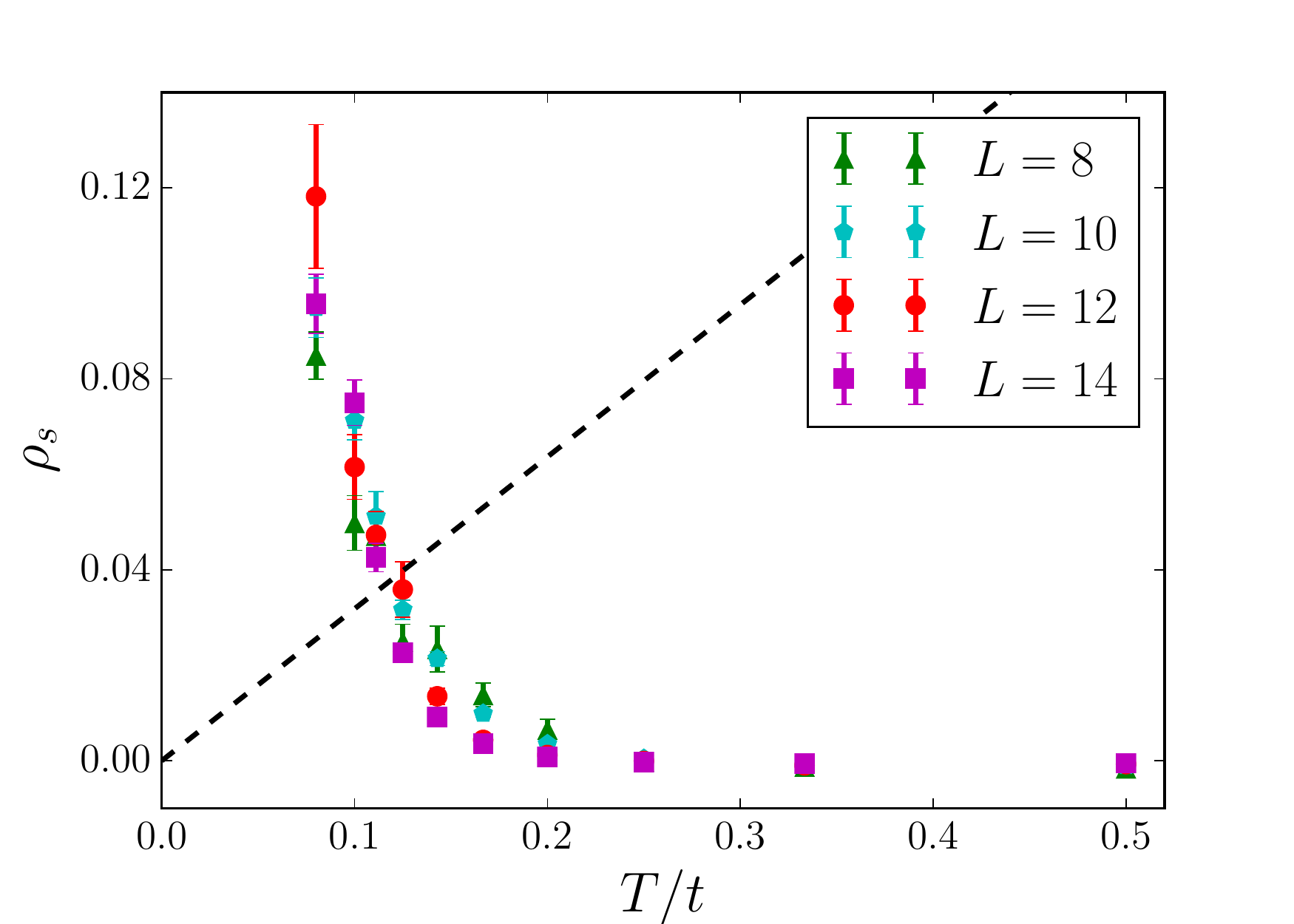}\caption{\label{fig:sfdensity}Superfluid density $\rho_{s}(L,T)$ as function
of temperature $T$ for the band dispersion $\delta/t=0.6$ and coupling
constant $\lambda^{2}=8t$ for various system sizes $L$. The BKT
transition temperature for each system size is determined by the condition
$\rho_{s}(L,T_{\mathrm{c}})=\frac{2}{\pi}T_{c}$. }
\end{figure}

Because the system is two-dimensional, the superconducting phase transition
is of the BKT type. Therefore, to determine $T_{\mathrm{c}}$, we
search for the temperature where the BKT condition is satisfied: 
\[
\rho_{s}(T_{\mathrm{c}})=\frac{2}{\pi}T_{c}
\]
where $\rho_{s}$ is the superfluid density. As explained in Ref.
\cite{Schattner16}, the latter can be extracted from our QMC simulations
via the current-current correlation function $\Lambda_{ij}$ according
to $\rho_{s}\equiv\lim_{L\rightarrow\infty}\rho_{s}\left(L\right)$,
with:
\begin{align}
\rho_{s}\left(L\right) & =\frac{1}{8}\sum_{a=x,y}\langle\Lambda_{aa}(q_{a}=\frac{2\pi}{L},q_{\bar{a}}=0,i\Omega_{n}=0)\rangle\\
 & -\frac{1}{8}\sum_{a=x,y}\langle\Lambda_{aa}(q_{a}=0,q_{\bar{a}}=\frac{2\pi}{L},i\Omega_{n}=0)\rangle
\end{align}
where $\bar{a}=y,x$ when $a=x,y$ and:

\begin{equation}
\Lambda_{ij}(\mathbf{r},\tau)\equiv\frac{1}{\beta L^{2}}\langle\int d\tau_{1}\sum_{\mathbf{r}_{1}}j_{i}(\mathbf{r}+\mathbf{r_{1},\tau+\tau_{1})}j_{j}(\mathbf{r}_{1},\tau_{1})\rangle
\end{equation}
with $j_{i}$ denoting the standard current operator. Note that the
model studied here is symmetric under the combination of a $\pi/2$
rotation, a particle-hole transformation, and the exchange of the
two bands, implying $\Lambda_{xx}(\mathbf{r},\tau)=\Lambda_{yy}(\tilde{\mathbf{r}},\tau)$,
where $\mathbf{r}$ and $\tilde{\mathbf{r}}$ are related by a $\pi/2$
rotation. Fig. \ref{fig:sfdensity} shows $\rho_{s}$ for various
system sizes for the band dispersion $\delta/t=0.6$ and the interaction
parameter $\lambda^{2}=8t$. The estimated transition temperature
$T_{c}(L)$ for each system of size $L$ is determined as the intersection
between the interpolated curve of $\rho_{s}(L,T)$ and $\frac{2}{\pi}T$.
The error bars in $\rho_{s}$ arising from the QMC sampling are used
to estimate the error bars of $T_{c}$ in the following way: besides
the interpolation curve passing through the average values of $\rho_{s}$,
we also determine two additional interpolation curves passing through
the top and the bottom of each error bar related to $\rho_{s}$. The
error bars in $T_{c}$ are estimated by determining when these two
additional curves cross $\frac{2}{\pi}T$.

\subsection{Thermodynamic limit of the BKT transition temperature}

\begin{figure}
\begin{centering}
\includegraphics[width=0.7\columnwidth]{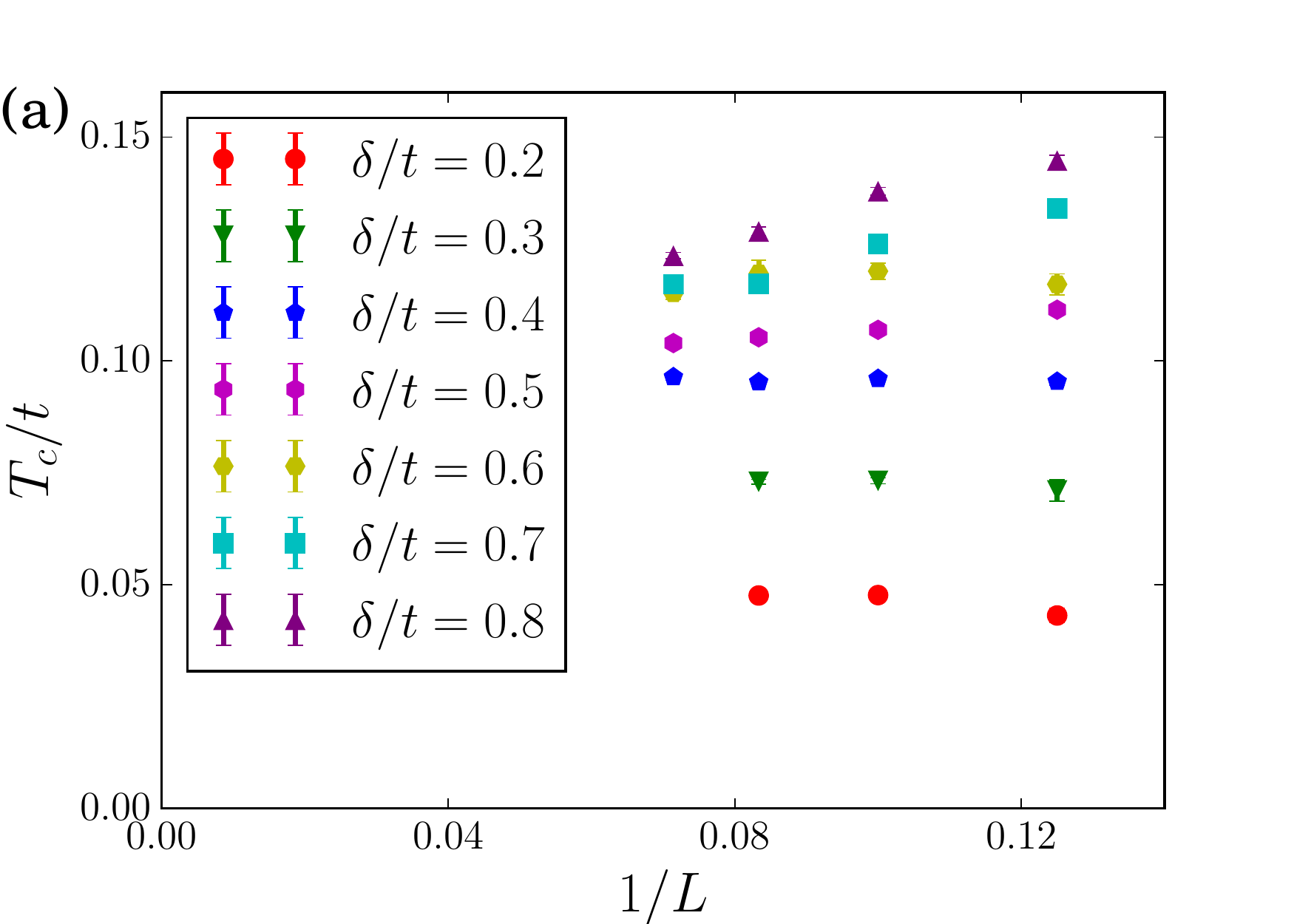}
\par\end{centering}
\begin{centering}
\bigskip{}
\par\end{centering}
\centering{}\includegraphics[width=0.45\columnwidth]{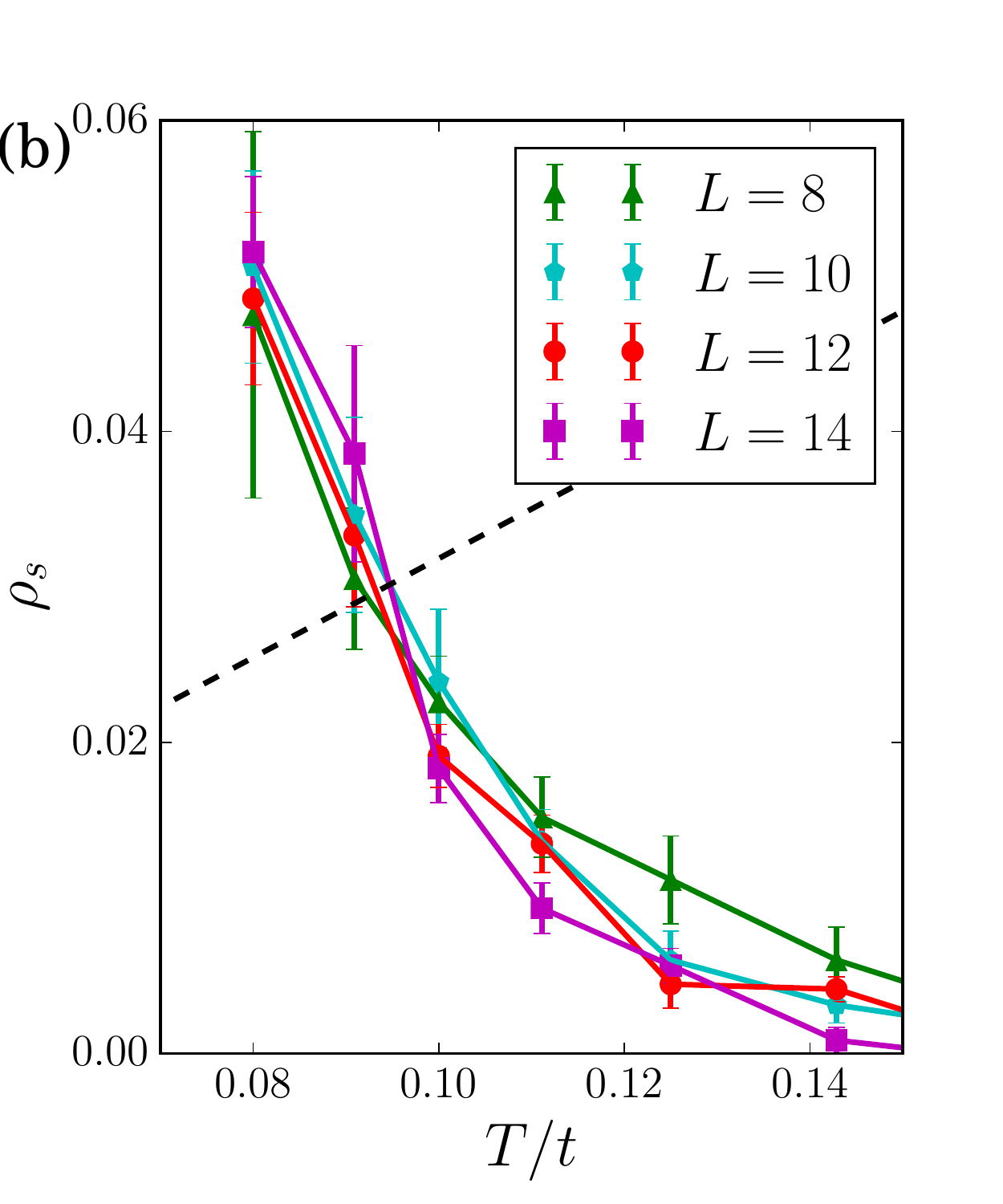}\includegraphics[width=0.45\columnwidth]{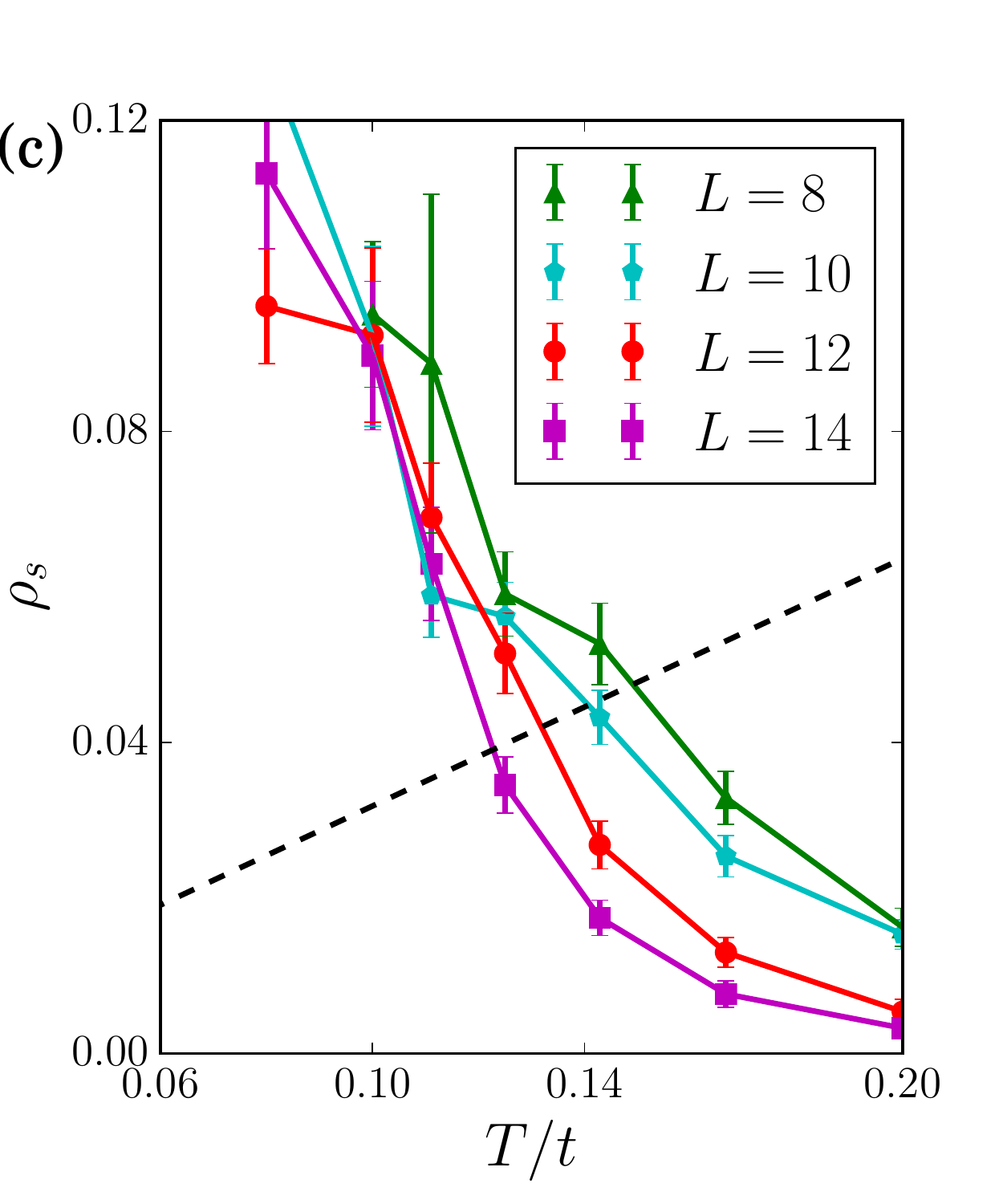}\caption{\label{fig:BKT_thermo}(a) The QMC extracted $T_{c}\left(L\right)$
as function of the inverse system size $1/L$ for all band dispersion
parameters $\delta/t$. Interpolated $\rho_{S}(T)$ curve for $\delta/t=0.4$
(b) and for $\delta/t=0.8$ (c).}
\end{figure}

To estimate the thermodynamic value of the BKT transition, we first
plot the extracted $T_{c}\left(L\right)$ as function of $1/L$ in
Fig. \ref{fig:BKT_thermo}a. For most of the values of $\delta/t$
that we studied \textendash{} specifically, $\delta/t=0.2,\,0.3,\,0.4,\,0.5,\,0.7$
\textendash{} we found a near saturation of $T_{c}(L)$ for the two
largest system sizes studied, i.e. $L=12$ and $L=14$ for $0.4\leq\delta/t\leq0.8$,
and $L=10$ and $L=12$ for $0.2\leq\delta/t\leq0.3$. We verified
that the reason for this behavior is that the superfluid density curves
for the two largest system sizes agree within statistical error bars
near the BKT transition. We illustrate this behavior for the case
$\delta/t=0.4$ in Fig. \ref{fig:BKT_thermo}b. Therefore, for these
band dispersions, we estimate the thermodynamic value for the transition
temperature to be given by $T_{c}\left(L_{\mathrm{max}}\right)$.

For the band dispersion with $\delta/t=0.6$, even though $T_{c}$
nearly saturates for the two largest system sizes, the corresponding
superfluid density curves are not on top of each other within the
QMC statistical error bars. This is also the case for the band dispersion
with $\delta/t=0.8$, as shown in Fig. \ref{fig:BKT_thermo}c. Moreover,
for this band dispersion, $T_{c}$ does not really seem to saturate
for the two largest system sizes, as shown in Fig. \ref{fig:BKT_thermo}a.
For these two systems, $T_{c}\left(L_{\mathrm{max}}\right)$ should
therefore be understood as an upper bound value for the thermodynamic
value of $T_{c}$. In these cases, we can also estimate the lower
bound value by the condition that the $\rho_{s}(T,L_{\mathrm{max}})$
curve becomes larger than $\rho_{s}(T,L)$ for one of the smaller
system sizes studied (in our case, $L=12$). Such a criterion is based
on the fact that, in the disordered phase, finite-size effects generally
make $\rho_{s}(T,L)$ larger for smaller system sizes. The extracted
lower boundary values for $T_{c}$ are depicted as the stars in Fig.
3 of the main text. Clearly, the only system where finite size effects
are more pronounced is the one with $\delta/t=0.8$.

\subsection{Density of states of the finite-size system}

\begin{figure}
\includegraphics[width=0.8\columnwidth]{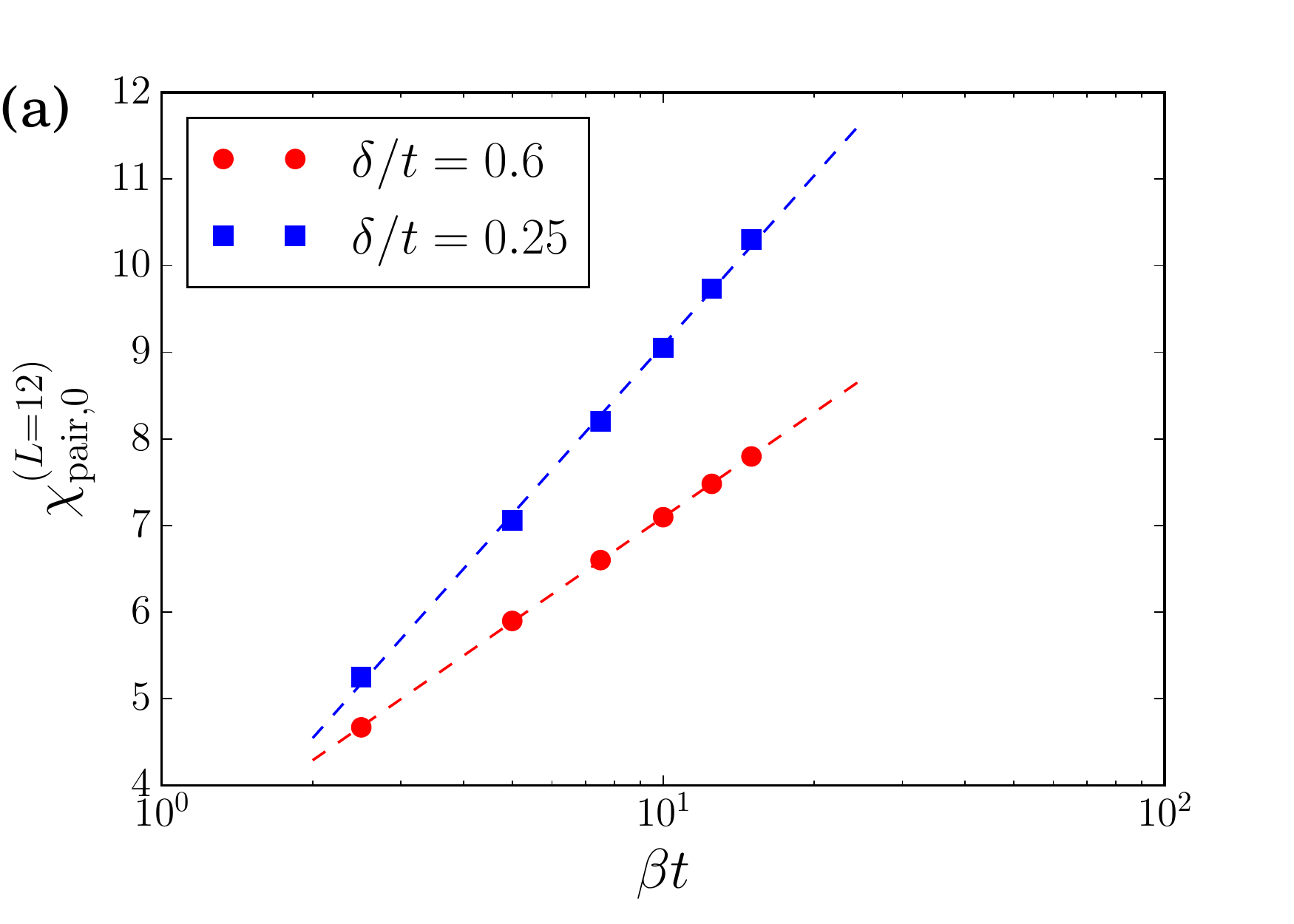}
\includegraphics[width=0.8\columnwidth]{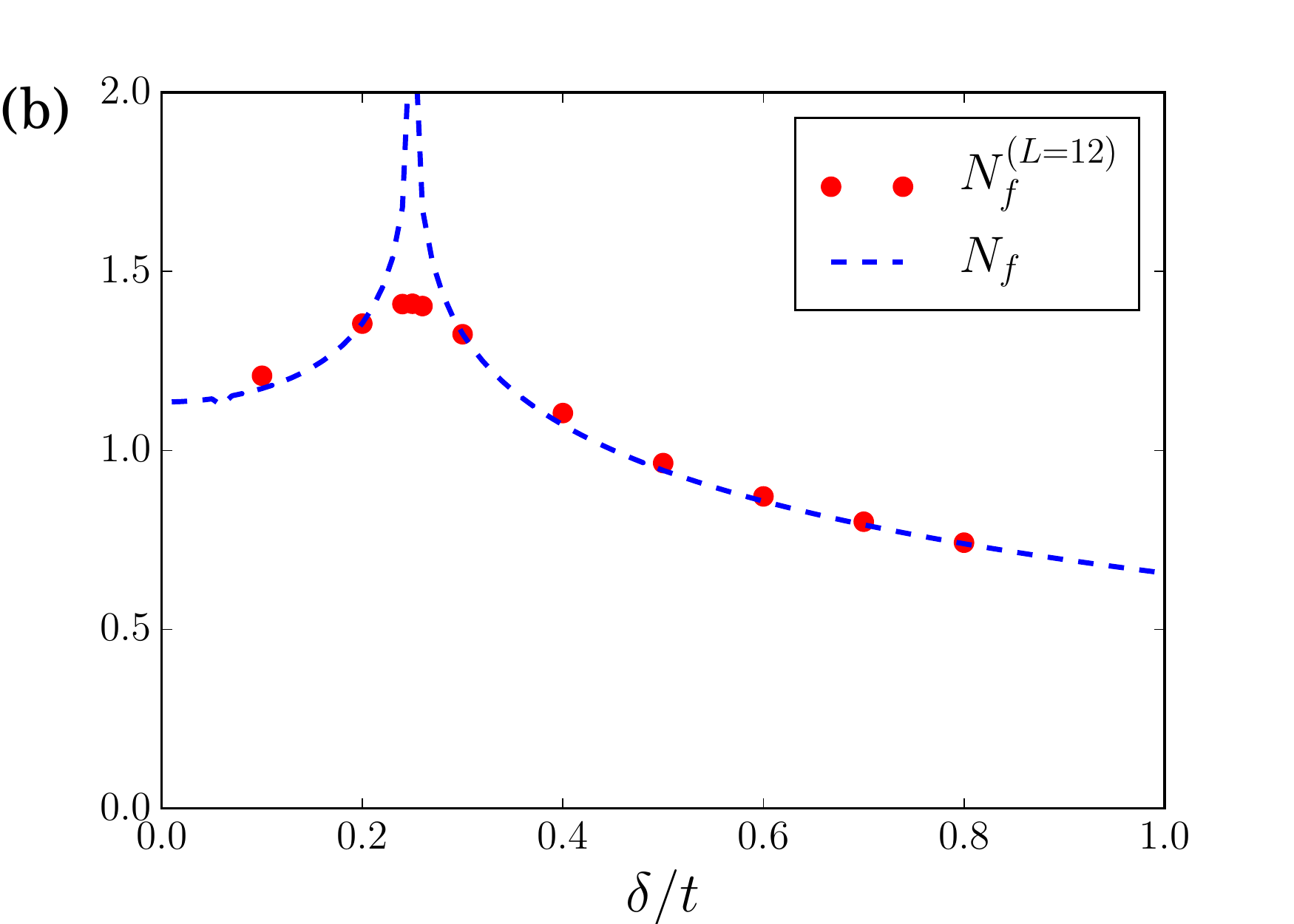}
\caption{\label{fig:dos}(a) $\chi_{\mathrm{pair},0}^{(L=12)}$ of the finite-size
system $L=12$ plotted as a function of the inverse temperature $\beta$
(in units of $1/t$) at the van Hove point ($\delta/t=0.25$) and
at $\delta/t=0.6$. (b) Comparison between the density of states of
the finite-size system, $N_{f}^{(L=12)}$, and the density of states
computed analytically, $N_{f}$. The results match except very close
to the van Hove singularity, where the divergence is cut off by finite
size effects. }
\end{figure}

Here we demonstrate that the bare pairing susceptibility $\chi_{\mathrm{pair,0}}$
in our simulations is sensitive to the proximity to the van Hove singularity,
despite the modest sizes of the systems. From Eq. \ref{eq:non_int_pair_sus},
we have:

\begin{equation}
\chi_{\mathrm{pair},0}(\beta)=2N_{f}\ln\Lambda\beta
\end{equation}
where $\Lambda$ is an upper cutoff related to the band structure.
In Fig. \ref{fig:dos}a, we show the exactly calculated pairing susceptibility
for a system of size $L=12$, $\chi_{\mathrm{pair},0}^{(L=12)}$.
We also show linear fittings to the expression above, from which we
can extract the density of states of the finite-size system, $N_{f}^{(L=12)}$.
In Fig. \ref{fig:dos}b, we compare $N_{f}^{(L=12)}$ to the analytically
calculated $N_{f}=-\frac{4}{\pi}\sum_{\mathbf{k}}\lim_{\delta\rightarrow0_{+}}G(\mathbf{k},i\delta)$
as function of $\delta/t$. The factor of $4$ arises from spin and
band degeneracies. The agreement between $N_{f}^{(L=12)}$ and $N_{f}$
is evident, and the only effect of the finite size of the system is
to cut-off the divergence of $N_{f}$ at the van Hove point.

\subsection{Behavior of $T_{c}$ for larger interaction parameters}

\begin{figure}
\includegraphics[width=0.9\columnwidth]{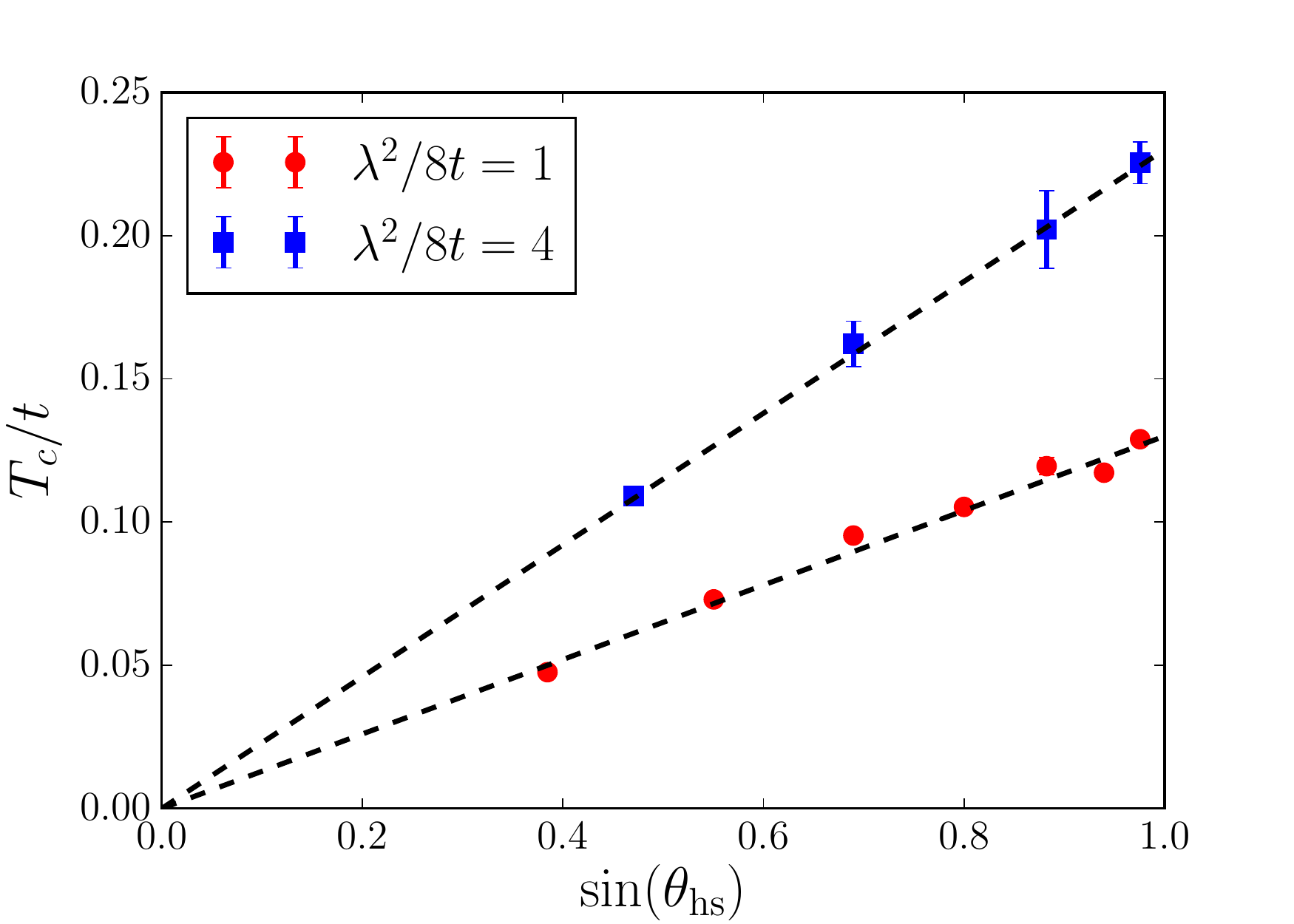}\caption{\label{fig:large_yukawa}$T_{c}$ as a function of $\sin\left(\theta_{\mathrm{hs}}\right)$
for $\lambda^{2}/8t=1$, where $T_{c}$ scales with $\lambda^{2}$,
and $\lambda^{2}/8t=4$, where $T_{c}$ is in the saturation regime
(see Fig. 5 of the main text). The linear dependence of $T_{c}$ on
$\sin\theta_{\mathrm{hs}}$ remains robust for larger values of the
interaction parameter $\lambda$. The results in this figure are obtained
for $L=12$.}
\end{figure}

To complement the discussions in Fig. 5 of the main text, in Fig.
\ref{fig:large_yukawa} we present $T_{c}$ as a function of $\sin\theta_{\mathrm{hs}}$
for both $\lambda^{2}/8t=1$ and $\lambda^{2}/8t=4$ \textendash{}
which is the largest interaction parameter studied. For the latter,
$T_{c}$ is in the saturation regime, as shown in Fig. 5 of the main
text. Note, however, that $T_{c}$ is still linearly proportional
to $\sin\theta_{\mathrm{hs}}$, without any enhancements due to the
van Hove singularity at $\delta/t=0.25$. Interestingly, the extension
of the hot-spot Eliashberg calculation discussed in Eq. (\ref{eq:eliash_eqnv2})
still predicts a linear dependence of $T_{c}$ with $\sin\theta_{\mathrm{hs}}$
even when $T_{c}$ becomes independent on $\lambda$.

\end{document}